\documentclass[aps,english,prl,amsmath,amsfonts,amssymb,superscriptaddress,twocolumn,showpacs,citeautoscript]{revtex4-1}
\usepackage[T1]{fontenc}
\usepackage[latin9]{inputenc}
\usepackage{amsmath}
\usepackage{amssymb}
\usepackage{wasysym}
\usepackage{esint}
\usepackage{graphicx}
\usepackage{times}
\usepackage{nicefrac}
\usepackage{appendix}
\usepackage{textcase}
\usepackage{subfigure}
\usepackage{empheq}
\usepackage{color}
\usepackage{leftidx}
\usepackage{makecell}
\usepackage{stackrel}
\makeatletter
\@ifundefined{textcolor}{}
{%
 \definecolor{BLACK}{gray}{0}
 \definecolor{WHITE}{gray}{1}
 \definecolor{RED}{rgb}{1,0,0}
 \definecolor{GREEN}{rgb}{0,1,0}
 \definecolor{BLUE}{rgb}{0,0,1}
 \definecolor{CYAN}{cmyk}{1,0,0,0}
 \definecolor{MAGENTA}{cmyk}{0,1,0,0}
 \definecolor{YELLOW}{cmyk}{0,0,1,0}
}

\renewcommand{\vec}[1]{\mathbf{#1}}
\renewcommand{\Re}{\operatorname{Re}}
\renewcommand{\Im}{\operatorname{Im}}

\newcommand{\asym}{\operatorname{asym}}

\renewcommand{\b}{\beta}

\newcommand{\add}[1]{\if\a\b{{\color{red} #1}}\else{#1}\fi}

\newcommand{\bracket}[1]{\langle #1 \rangle}
\newcommand{\ket}[1]{| #1 \rangle}
\newcommand{\bra}[1]{\langle #1 |}
\newcommand{\im}{\mathrm{i}}

\renewcommand{\eqref}[1]{(\ref{eq:#1})}

\newcommand{\Figref}[1]{Figure~\ref{fig:#1}}

\newcommand{\trace}[1]{{\rm Tr} \left[ #1 \right]}

\newcommand{\Tr}{\text{Tr }}

\makeatother
\usepackage{babel}

\begin{document}
\title{Phonon-polariton mediated thermal radiation and heat transfer
  among molecules and macroscopic bodies: nonlocal electromagnetic
  response at mesoscopic scales}

\author{Prashanth S. Venkataram}
\affiliation{Department of Electrical Engineering, Princeton University, Princeton, New Jersey 08544, USA}
\author{Jan Hermann}
\affiliation{Physics and Materials Science Research Unit, University of Luxembourg, L-1511 Luxembourg}
\author{Alexandre Tkatchenko}
\affiliation{Physics and Materials Science Research Unit, University of Luxembourg, L-1511 Luxembourg}
\author{Alejandro W. Rodriguez}
\affiliation{Department of Electrical Engineering, Princeton University, Princeton, New Jersey 08544, USA}

\date{\today}

\begin{abstract}
  Thermal radiative phenomena can be strongly influenced by the
  coupling of phonons and long-range electromagnetic fields at
  infrared frequencies. Typically employed macroscopic descriptions of
  thermal fluctuations tend to ignore atomistic effects that become
  relevant at nanometric scales,
  whereas purely microscopic treatments ignore long-range,
  geometry-dependent electromagnetic effects. We describe a mesoscopic
  framework for modeling thermal fluctuation phenomena among molecules
  in the vicinity of macroscopic bodies, conjoining atomistic
  treatments of electronic and vibrational fluctuations obtained from
  ab-initio density functional theory in the former with continuum
  descriptions of electromagnetic scattering in the latter.  The
  interplay of these effects becomes particularly important at
  mesoscopic scales, where phonon polaritons can be strongly
  influenced by the finite sizes, shapes, and non-local/many-body
  response of the bodies to electromagnetic fluctuations. We show that
  even in small but especially in elongated low-dimensional molecular
  systems, such effects can modify thermal emission and heat transfer
  by orders of magnitude and produce qualitatively different behavior
  compared to predictions based on local, dipolar, or pairwise
  approximations valid only in dilute media.

\end{abstract}

\maketitle

Radiative heat transfer plays an important role in the
characterization of molecular structures, including graphene,
fullerenes, carbon nanotubes, and other low-dimensional materials, and
has applications to thermophotovoltaic power
generation~\cite{MessinaNATURE2013, SvetovoyPRAPP2014, IlicOE2012,
  LenertNATURENANO2014}, lasers~\cite{RuanPRB2006,
  BoriskinaNANOPHOT2016}, and single-molecule
junctions~\cite{CuiJCP2017, CuiSCIENCE2017}. Recent
theoretical~\cite{CuiJPCA2015, PendryPRB2016, EdalatpourPRB2016} and
experimental~\cite{ChiloyanNATURE2015, KloppstechNATURE2017,
  CuiNATURE2017, StGelaisNANOLETT2014} works in this area have sought
to explore transport at nanometric scales, where the interplay of
electromagnetic (EM) and phonon-mediated transport can no longer be
captured solely by macroscopic
electrodynamics~\cite{ChiloyanNATURE2015, KloppstechNATURE2017,
  CuiNATURE2017}. We present a framework for modeling thermal
fluctuation phenomena among molecules in the vicinity of continuum
bodies at mesoscopic scales and apply it to compute heat transfer
between molecules in contact with separate thermal reservoirs. Our
approach captures the emergence of phonon
polaritons~\cite{TianPRB2014, TianPRB2012}, which typically arise at
infrared frequencies and hence probe the bulk of the Planck
distribution near room temperature, and the impact on charge
delocalization (nonlocality) and radiation by: First, extending
microscopic ab-initio models of electronic response in molecular
bodies based on density functional theory (DFT)
to include molecular vibrations mediated by short-range interactions
(phonons); Second, combining this with classical EM scattering
techniques applicable to \emph{arbitrary} macroscopic
objects~\cite{RodriguezPRB2013, PolimeridisPRB2015} to account for
their coupling to collective electronic excitations (plasmon
polaritons), including retardation, multiple scattering, and
geometry-dependent many-body effects to all orders. Our framework,
detailed below, assumes a harmonic approximation to the chemical bonds
effecting molecular phonons and ignores thermally driven changes in
molecular conformations, both of which are only expected to be
violated at large temperatures. For illustration, we investigate
thermal emission and heat transfer among fullerene and carbyne
molecules at different temperatures, either in vacuum or above a
metallic surface. We find differences of over an order of magnitude in
the emitted and transferred powers relative to common dipolar
approximations that neglect finite size and many-body EM effects, with
the greatest differences obtained for elongated carbyne wires. We show
that the proximity of the molecules to the metallic surface and
resulting induction of image charges can qualitatively change their
mutual heat transfer. More interestingly, we find that molecular
phonon polaritons can delocalize the spatial extent of their
polarization response, from a few angstroms up to tens of nanometers,
in ways which depend on the shape of the molecules and their proximity
to the surface. Such non-locality causes the heat transfer to change
qualitatively (eventually saturating) at nanometric scales.

Theoretical models that account for atomic-scale and vibrational
effects on radiative emission and heat transfer exist, but suffer from
a number of limitations~\cite{HajianJOSAB2017, ShenNANOLETT2009,
  GarciaDeAbajoACSNANO2013, BasovSCIENCE2016,LuoPCCP2013,
  BoriskinaNANOPHOT2016, CareyNMTE2008, BoriskinaMT2013,
  CahillAPR2014}.
A common approach involves application of atomistic Green's
functions~\cite{DharJSP2006, TianPRB2014, TianPRB2012, MingoPRB2003,
  ChiloyanNATURE2015, PendryPRB2016} to model phonon-mediated
transport across material boundaries. While the harmonic spring
constants that lead to collective mechanical oscillations have
recently been derived through DFT methods, they have thus far only
been applied to bulk media, whereas heat transport across vacuum gaps
or in structured media has been largely ignored or approximated
through ad-hoc, pairwise models known to fail in complex
geometries~\cite{TkatchenkoJCP2013, GobreNCOMMS2013, DiStasioJPCM2014,
  TkatchenkoADFM2014, AmbrosettiSCIENCE2016}. Another class of
atomistic methods involves molecular dynamics~\cite{HenryPRL2008,
  EsfarjaniPRB2011, NoyaPRB2004, CuiJPCA2015},
which capture the anharmonicity of covalent bonds but suffer from
similar pitfalls of heuristically approximating noncovalent
interactions. Other approaches are based on continuum
fluctuational electrodynamics. These include discrete dipolar and
multipolar methods~\cite{MessinaPRB2013, PerezMadridPRB2008,
  EdalatpourJQSRT2014, EdalatpourPRB2016, EdalatpourJQSRT2016}, which
approximate point dipolar responses as arising from bulk
susceptibilities, as well as methods based on solutions of Maxwell's
equations in continuum media~\cite{YuNATURE2017, KrugerPRB2012,
  MoleskyPRB2015, ZhaoPRB2017, MessinaNATURE2013, SvetovoyPRAPP2014,
  IlicOE2012,RodriguezPRB2013, PolimeridisPRB2015, OteyJQSRT2014,
  ReidPROCIEEE2013, RodriguezPRL2011, LuoPRL2004}. While macroscopic
electrodynamic treatments can capture the impact of long-range EM
effects, they typically treat material responses at atomic scales via
semi-empirical models rather than through ab-initio calculations, and
can therefore miss important effects arising from the spatial
dispersion or nonlocality of electronic wavefunctions and its
interplay with geometry.
Even for a material like graphene with highly delocalized electronic
and mechanical response~\cite{JablanPRB2011, HwangPRB2010,
  XiaoFOP2016, SvetovoyPRAPP2014}, nonlocality is typically treated
phenomenologically through macroscopic approximations of quantum
behavior. Motivated by recent predictions of~\cite{GobreNCOMMS2013,
  DiStasioJPCM2014, TkatchenkoADFM2014, AmbrosettiSCIENCE2016,
  VenkataramPRL2017} of many-body van der Waals (vdW) interactions in
low-dimensional systems and experiments of heat transfer at the
nanoscale~\cite{ChiloyanNATURE2015, KloppstechNATURE2017,
  CuiNATURE2017, StGelaisNANOLETT2014, SongNATURENANO2015}, both of
which are starting to probe regimes where continuum electrodynamics is
no longer valid,
we propose a 
framework that can capture such thermal
radiative processes at mesoscopic scales.




\emph{Background.---} We consider a collection of $N_\mathrm{mol}$
molecules labeled $k$ and described by electric susceptibilities
$\mathbb{V}_{k}$, relating induced polarization and electric
fields. For simplicity, we consider molecules which are maintained at
temperatures $T_{k}$ in the presence of a collection of macroscopic
bodies, described collectively by a macroscopic electric
susceptibility $\mathbb{V}_{\mathrm{env}}$ and maintained at a
different temperature $T_{\mathrm{env}}$ (henceforth assumed to be
zero for conceptual convenience). The radiative energy transfer from
molecule $m$ to $n$ (which may or may not be the same), can be shown
to be~[see appendix]:
\begin{equation} \label{eq:Phimn}
  \Phi_{n}^{(m)} = -\frac{1}{2\pi}
  \trace{\Im(\mathbb{V}_{m}^{-1\star}) \mathbb{P}_{m}
    \mathbb{T}^{\star} \asym\left(\mathbb{G}_{\mathrm{env}}
    \mathbb{P}_{n}\right) \mathbb{T} \mathbb{P}_{m}},
\end{equation}
where
$\mathbb{T}^{-1} = \sum_{k} \mathbb{V}_{k}^{-1} -
\mathbb{G}_{\mathrm{env}}$ is known as a scattering transition
operator, which describes EM scattering (to all orders) by the
collection of molecules in the presence of all
bodies~\cite{RahiPRD2009,KrugerPRB2012},
$\mathbb{G}_{\mathrm{env}} = (\mathbb{G}_{0}^{-1} -
\mathbb{V}_{\mathrm{env}})^{-1}$ is the classical electromagnetic
Green's function~\cite{RahiPRD2009,KrugerPRB2012,VenkataramPRL2017},
and $\mathbb{P}_{k}$ is a projection onto the space spanned by the
degrees of freedom of molecule $k$. Here, ${}^{\star}$ implies complex
conjugation, ${}^{\dagger}$ implies conjugate transposition (adjoint),
$\Im(\mathbb{A}) = \frac{\mathbb{A} - \mathbb{A}^{\star}}{2i}$, and
$\asym(\mathbb{A}) = \frac{\mathbb{A} -
  \mathbb{A}^{\dagger}}{2i}$. Given \eqref{Phimn}, the far-field
thermal emission from molecule $m$,
\begin{equation}
  W^{(m)} = \sum_{n = 1}^{N} s_{nm} \Phi_{n}^{(m)} \Theta(\omega,
  T_{n}),
\label{eq:emission}
\end{equation}
and the net heat transfer from molecule $m$ to molecule $n$,
\begin{equation}
  W_{m \to n} = \Phi_{n}^{(m)} [\Theta(\omega, T_{m}) - \Theta(\omega,
    T_{n})],
\label{eq:heat}
\end{equation}
where $\Theta(\omega,T)$ is the Planck function while the sign
function $s_{nm} = 1 - 2\delta_{nm}$ accounts for the direction of
energy flow for the emission of a body. In either case, the net power
is the integral over all frequencies,
$P = \int_{0}^{\infty} W(\omega)~\mathrm{d}\omega$.

The above formulas require accurate and fast computations of
$\mathbb{G}_{\mathrm{env}}$ and $\mathbb{V}_{k}$. The former can be
computed using any number of state-of-the-art classical EM
techniques~\cite{RodriguezPRB2013, PolimeridisPRB2015, OteyJQSRT2014,
  ReidPROCIEEE2013, RodriguezPRL2011, LuoPRL2004}. The molecular
susceptibilities $\mathbb{V}_{k}$ generally require quantum
descriptions, but recent work in the related field of vdW
interactions~\cite{TkatchenkoJCP2013, DiStasioJPCM2014,
  TkatchenkoADFM2014, AmbrosettiJCP2014} has shown that accurate
models of the valence electronic response of insulating or weakly
metallic molecules, including carbon allotropes, can be obtained by
expressing
$\mathbb{V}_{k} = \sum_{p, q} \alpha_{pi,qj} \ket{f_{p} \vec{e}_{i}}
\bra{f_{q} \vec{e}_{j}}$, in a localized basis $\ket{f_{p}}$ (along
Cartesian direction $\vec{e}_{i}$) associated with each atom $p$ by
modeling its valence electrons as a quantum oscillator in its ground
state. We extend that framework in the following way: each atom $p$ in
a given molecule is comprised of an effective valence electron which
couples to long-range EM fields and has charge $q_{p}$, mass
$m_{\mathrm{e} p}$, and damping coefficient
$b_{\mathrm{e} p} = m_{\mathrm{e} p} \gamma_{\mathrm{e} p}$, as well
as a nucleus of mass $m_{\mathrm{I} p}$ which is coupled only to its
associated valence electron with spring constant $k_{\mathrm{e} p}$
and to neighboring nuclei~\cite{TianPRB2014, TianPRB2012, RuanPRB2006,
  CareyNMTE2008} with spring constants $\mathbb{K}_{pq}$. While the
nuclear masses are obtained from elemental data and the electronic
oscillator damping coefficients $\gamma_{\mathrm{e}p}$ from empirical
data, the masses, charges, and spring constants are all obtained from
DFT~\cite{DiStasioJPCM2014, AmbrosettiSCIENCE2016}, thereby accounting
for short-range electrostatic, hybridization, and quantum exchange
effects; this includes the nuclear harmonic couplings from
lowest-order expansions of ab-initio DFT force-constant computations,
which leads to anisotropy (tensorial couplings) due to the
directionality of covalent bonds. This leads to the simple
frequency-domain equations of motion:
\begin{equation}
  \begin{bmatrix}
  K_{\mathrm{e}} - \im\omega B_{\mathrm{e}} - \omega^{2} M_{\mathrm{e}} & - K_{\mathrm{e}}  \\
- K_{\mathrm{e}} &   K_{\mathrm{e}} + K_{\mathrm{I}} - \omega^{2} M_{\mathrm{I}}
\end{bmatrix}
\begin{bmatrix}
  x_{\mathrm{e}} \\
  x_{\mathrm{I}}
\end{bmatrix}
= 
\begin{bmatrix}
  Q_{\mathrm{e}} e_{\mathrm{e}} \\
  0
\end{bmatrix},
\end{equation}
where $(Q_{\mathrm{e}}, M_{\mathrm{e}}, M_{\mathrm{I}},
K_{\mathrm{e}}, K_{\mathrm{I}}, B_{\mathrm{e}})$ are $3N_{k} \times
3N_{k}$ matrices representing, respectively, the charges and masses of
the electronic oscillators, the masses of the nuclei, the spring
constants binding the electronic oscillators to the respective nuclei,
the internuclear harmonic couplings, and the damping coefficients of
the electronic oscillators. The latter determine the amplitude
response of the nuclear positions $x_{\mathrm{I}}$ and electronic
dipole moments $p_{\mathrm{e}} = Q_{\mathrm{e}} x_{\mathrm{e}}$ in the
presence of an electric field, represented by a $3N_{k}$-dimensional
vector $e_{\mathrm{e}}$ obtained by evaluating $\ket{\vec{E}}$ at
every atomic position. Solving for $p_{\mathrm{e}} = \alpha
e_{\mathrm{e}}$ yields the electric susceptibility matrix,
\begin{equation}
  \alpha = Q_{\mathrm{e}} (K_{\mathrm{e}} - \im\omega B_{\mathrm{e}} -
  \omega^{2} M_{\mathrm{e}} - K_{\mathrm{e}} (K_{\mathrm{e}} +
  K_{\mathrm{I}} - \omega^{2} M_{\mathrm{I}})^{-1}
  K_{\mathrm{e}})^{-1} Q_{\mathrm{e}},
\end{equation}
which enters the expansion of $\mathbb{V}_{k}$ above. The discrepancy
in the mass of electronic and ionic oscillators
ensures that the poles of $\alpha$ separate into ultraviolet
resonances, corresponding to predominantly electronic modes, and
infrared resonances, corresponding to phononic modes, the latter of
which is primarily responsible for thermal emission.

The form of $K_{\mathrm{I}}$, coupling neighboring nuclei, ensures
that nonlocality enters the polarization response in a microscopic
rather than phenomenological manner. Hence, in contrast to point
dipoles or macroscopic objects modeled via local
susceptibilities~\cite{VolokitinPRB2001, DominguesPRL2005,
  ShenNANOLETT2009, CuiNATURE2017, KloppstechNATURE2017}, which lead
to diverging EM fields as two or more bodies approach one another, the
molecular susceptibility above will always be finite. In particular,
the electronic response is accurately described by localized Gaussian
basis functions~\cite{DonchevJCP2006, AmbrosettiSCIENCE2016,
  DiStasioJPCM2014, PhanJAP2013, ShtogunJPCL2010, KimLANGMUIR2007,
  ColeMS2009},
\begin{equation}
  f_{p} (\vec{x}) = \left(\sqrt{2\pi}
    \sigma_{p} \right)^{-3}
  \mathrm{exp}\left(-\frac{ (\vec{x} - \vec{x}_{p})^2 }{2\
      \sigma_{p}^{2}}\right),
\end{equation}
centered at the locations $\vec{x}_{p}$ of each atom $p$, normalized
such that $\int \mathrm{d}^{3} \vec{x}~f_{p} = 1$, and featuring a
width that, rather than being phenomenological~\cite{MahantyJCSFT1975,
  RennePHYSICA1971A}, depends on the bare atomic polarizability at
every frequency via
$\sigma_{p} (\omega) = \left(\frac{|\alpha_{p}
    (\omega)|}{\sqrt{72\pi^3}}\right)^{1/3}$~\cite{MayerPRB2007,
  DiStasioJPCM2014} where
$\alpha_{p} (\omega) = \sum_{q,j} \alpha_{pj,qj} (\omega)$. Such basis
functions mitigate short-distance EM divergences through the smearing
of the charge density over non-zero widths $\sigma_{p}$, leading to
finite matrix elements,
$\bracket{f_{p} | (\cdots) f_{q}} = \int \mathrm{d}^{3}
\vec{x}~\mathrm{d}^{3} \vec{x}'~f_{p} (\vec{x}) (\cdots) f_{q}
(\vec{x}') < \infty$, even for coincident atoms $p = q$. While these
widths are typically smaller than 1
angstrom~\cite{AmbrosettiSCIENCE2016, DiStasioJPCM2014}, we find that
the introduction and delocalized nature of phonons smears the
effective atomic polarization response in a way that depends strongly
on the shapes of the molecules, leading to dramatically larger widths,
on a scale $\xi \sim 1~\mathrm{nm}$. As we show below, this stronger
nonlocality has a noticeable impact on heat transfer among molecules
at nanometric separations. Furthermore, since $\alpha$ enters
$\mathbb{T}$, which includes long-range EM scattering, thermal
fluctuations will be largest near the frequency poles of $\mathbb{T}$,
corresponding to the dressed, \emph{phonon}-\emph{polariton}
resonances of the fully coupled hybrid molecular--macroscopic system.

\begin{figure}[t!]
\centering
\includegraphics[width=0.9\columnwidth]{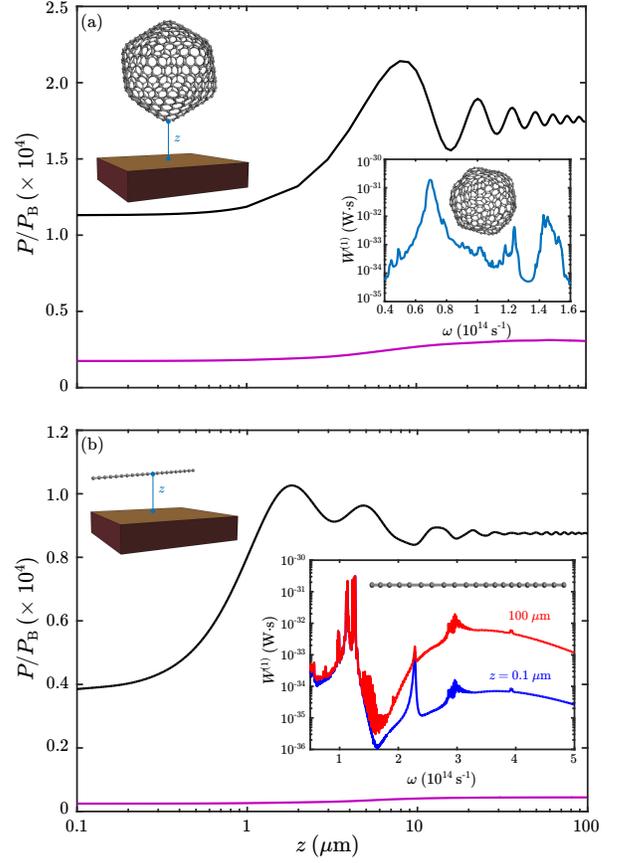}
\caption{\textbf{Contribution of phonon polaritons to molecular
    thermal emission.} (a) RMB (black) and DA (magenta) emission rates
  of a $\mathrm{C}_{500}$-fullerene held at $T=300$~K above a
  zero-temperature gold plate as a function of vertical separation
  $z$, normalized by the power emitted by a blackbody of the same area
  $P_\mathrm{B} = \sigma_{\mathrm{B}} T^{4} A$, where $A=4\pi R^2$ and
  $R=1$~nm.  Insets show a schematic of the fullerene above the plate
  as well as the RMB emission spectrum in the limit $z \to \infty$,
  along with a schematic of the dominant polaritonic mode. (b) Same as
  (a) but for a carbyne wire parallel to the plate, where
  $A = 2\pi RL$, $R = 0.13$~nm, and $L = 65~\mathrm{nm}$. The RMB
  emission spectrum is shown at both $z = 0.1~\mathrm{\mu m}$ (blue)
  and $z = 100~\mathrm{\mu m}$ (red).}
\label{fig:emission}
\end{figure}

\emph{Predictions.---}We now exploit this framework to present new
proof-of-concept predictions of thermal emission and heat
transfer. For simplicity, we consider $\mathbb{G}_{\mathrm{env}}$ as
arising from the field response in either vacuum or above a gold
plate, modeled as a perfect electrically conducting
surface~\footnote{For computational convenience, and since gold is
  essentially perfectly reflecting in the infrared frequency regime
  where nonequilibrium energy transfer is most significant, we model
  the gold plate as a perfect electrically conducting surface}. We
consider either a $\mathrm{C}_{500}$-fullerene of radius $1$~nm or a
500 atom-long carbyne wire of length $65$~nm in various configurations
above the plane; the geometries of these molecules and their material
parameters
$(Q_{\mathrm{e}}, M_{\mathrm{e}}, M_{\mathrm{I}}, K_{\mathrm{e}},
K_{\mathrm{I}})$ are taken from DFT, while the damping coefficients
$B_{\mathrm{e}}$ are assigned phenomenologically so that the loss
rates
agree with empirical observations of similar carbon
allotropes~\cite{BuhmannPRA2012}. Where appropriate, we explicitly
compare results obtained by evaluating the retarded many-body (RMB)
transfer rates~\eqref{Phimn}, which capture the impact of multiple
scattering, molecular geometry, and EM screening to all orders, to
those obtained from a dipolar approximation (DA) that contracts the
bare molecular responses $\mathbb{V}$ into a point particle, with the
susceptibilities and screened responses replaced by contracted dipolar
susceptibilities
$(\boldsymbol{\alpha}^\infty)_{ij} = \sum_{p,q} \alpha_{pi,qj}$. For
convenience, the integrated powers are normalized to the emitted power
$P_{\mathrm{B}} = \sigma_{\mathrm{B}} T^{4} A$ of a blackbody of the
same temperature $T$ and area $A$. Within this approximation, the
thermal emission rate of a molecule can be approximated (to lowest
order in the scattering) by,
\begin{equation}
  \Phi^{(0)}_{\infty} = -\frac{1}{2\pi} \Tr
  [\Im \boldsymbol{\alpha}^{\infty}\cdot
  \Im \mathbb{G}_{\mathrm{env}}]
\label{eq:DA}
\end{equation}
For a molecule at some position $\vec{r}_{0} = z\vec{e}_{z}$ above the
plate, $\mathbb{G}_{\mathrm{env}}$ can be obtained from image theory;
it is a constant in the near field and in the far field ($z \gg
\omega/c$) given by
$\Im \mathbb{G}_{\mathrm{env}}(\omega,z,z) = 
\frac{\omega^{3}}{6\pi c^{3}} \mathbb{I} - \frac{\omega^{2}}{8\pi
  c^{2} z} \sin\left(\frac{2\omega z}{c}\right) (\mathbb{I} -
\vec{e}_{z} \otimes \vec{e}_{z}).
$


\Figref{emission} shows the far-field thermal emission of either
molecule at $T = 300~\mathrm{K}$ as a function of their distance $z$
above the gold plate at zero Kelvin, normalized by the emission rate
$P_\mathrm{B}$ (on the order of $10^{-14}$~W) of blackbodies of the
same effective areas. The $\mathrm{C}_{500}$-fullerene is a compact
molecule whose vibrational modes and resulting polaritons are limited
in extent by its small size. As a result, its emission spectrum
consists of a small set of resonances in a narrow frequency range and
is dominated by a single peak at
$\omega_0 \approx 7\times 10^{13}~\mathrm{rad/s}$, causing the
emission to scale as $z^{-1}\sin(2\omega_0 z/c)$ in the limit of
$z \gg \lambda_T = \frac{\hbar c}{k_{\mathrm{B}} T} \approx
7.6~\mathrm{\mu m}$, where $\lambda_T$ is the thermal wavelength. At
small $z$, the emission approaches a constant value that is attenuated
relative to vacuum by screening of the field response near the gold
surface
. Similar behavior is observed in the case of a carbyne wire parallel
to the surface, except that the elongated shape of the molecule
enables longer-range vibrational modes and stronger polarization
anisotropy, which further suppresses emission at small $z$. For both
molecules, the DA \eqref{DA} underestimates the far-field emission by
almost an order of magnitude for all $z$, due to the neglect of
long-range EM scattering and screening effects. More dramatically, for
the carbyne wire in the RMB treatment, as $z$ increases, a set of
closely-spaced infrared polaritonic resonances begin to contribute,
causing a beating pattern in the emission as a function of $z$. These
close resonances do not arise within the DA, which only captures the
impact of short-range EM interactions on the polarization response.




\begin{figure}[t!]
\centering
\includegraphics[width=0.9\columnwidth]{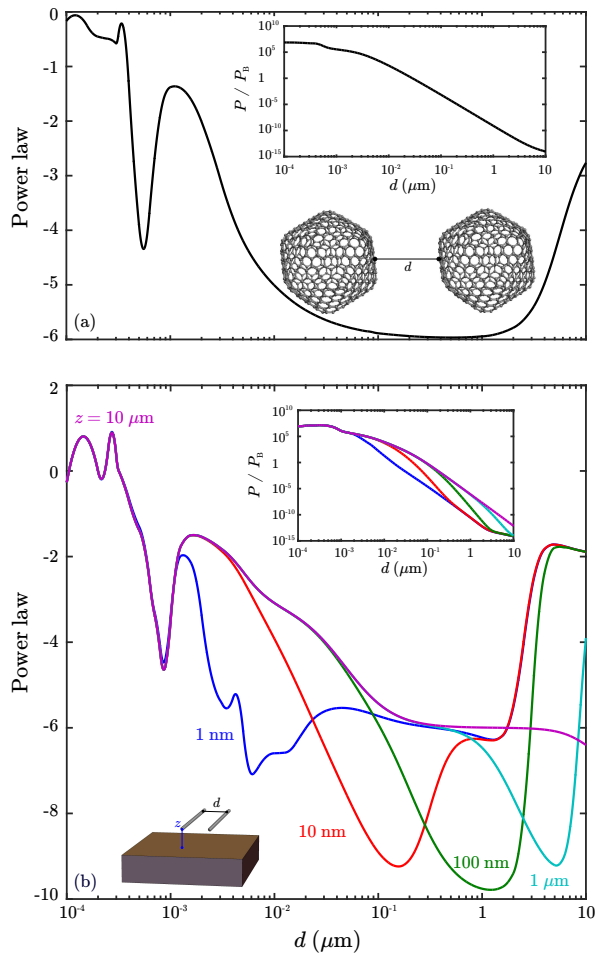}
\caption{\textbf{Impact of nonlocal response, shape, and many-body
    effects on heat exchange between molecules.} (a) Power-law of heat
  transfer, $\partial (\log P) / \partial (\log d)$, between two
  $\mathrm{C}_{500}$ fullerenes in vacuum held at $T=300$~K and zero
  Kelvin, with respect to their surface--surface separation $d$. The
  inset shows the transfer versus $d$, normalized by the emission rate
  from a corresponding, room-temperature blackbody $P_\mathrm{B}$. (b)
  Same as (a) but for two 500 atom-long parallel carbyne wires
  oriented parallel to a gold plate. The various curves denote
  different values of their vertical distance $z$ from the plate.}
\label{fig:heattransfer}
\end{figure}

\Figref{heattransfer} considers the heat transfer between two
molecules, one at $T = 300~\mathrm{K}$ and the other at zero
temperature, as a function of their mutual horizontal separation $d$
for several different values of their (identical) distances $z$ above
the zero-temperature plate. For two fullerenes far from the plate, we
find that the flux rate transitions from $1/d^2$ in the far field to
$1/d^{6}$ in the near field ($d \ll \lambda_T$), leading to flux rates
$P \gg P_\mathrm{B}$, consistent with the dipolar
approximation~\cite{VolokitinPRB2001, DominguesPRL2005}. However, as
$d$ decreases further to become comparable to the phonon-induced
nonlocal length scale $\xi = 1$~nm, the EM response begins to probe
the finite sizes and smeared electronic response of the molecules,
causing the heat transfer to saturate (in contrast to the diverging
flux rate arising in the dipolar picture) and the resulting power-law
exponent to approach 0. Hence, \eqref{heat} seamlessly captures the
transition between near-field radiation and conduction.




In the case of two parallel carbyne wires above a gold plate, the
departures from dipolar and pairwise predictions are even more
stark. For most separations, as $z$ decreases, the proximate surface
severely attenuates the heat flux relative to vacuum (by over 4 orders
of magnitude at $z \sim \xi$) except at very small values of
$d \lesssim \xi \ll z$ where the nonlocal response of the molecules
begins to saturate their response. In the far field $d \gg \lambda_T$,
as the molecules are much smaller than $\lambda_T$, the metal plate
does not affect the heat transfer much, so the power law attains the
far field dipolar limit of -2. As $d$ decreases further, if
$z \gg \lambda_T$, the power law reaches the near field dipolar limit
of -6, but if $z < \lambda_T$, the image dipoles induced in the
proximate surface lead to an effective triple-dipole configuration
with a power law of -9, akin to the so-called Axilrod--Teller--Muto
correction for dipolar vdW interactions~\cite{AxilrodJCP1943,
  DiStasioJPCM2014, TkatchenkoJCP2013}.
These power laws exhibit progressively stronger deviations from either
the dipolar limit of -6 or the pairwise limit of -5 (obtained by
summing the dipolar heat transfer between pairs of atoms across these
parallel linear molecules) in the near field as $d$ decreases further,
due to the confluence of various competing length scales, including
$z$, $d$, the wire lengths, $\xi$, and $\lambda_T$. If $z > \xi$, the
heat transfer begins to saturate for $d \lesssim 10$~nm due to charge
smearing. The situation becomes more complicated when both
$d,z \lesssim \xi$~\footnote{Note that in principle, as $d$ attains
  angstrom length scales, the assumption of the molecules being
  separate breaks down due to the possible formation of new covalent
  bonds; while this would require recomputing $K_{\mathrm{I}}$ for the
  coupled molecular system, our formalism for computing radiative heat
  transfer still holds as long as the two molecular components can be
  maintained at different temperatures.}  due to the interplay of
object sizes and nonlocality, with the finite and elongated
one-dimensional geometry of the parallel wires also playing a
significant role. In this regime, the nonlocal response of the wires
begins to have an effect on their images in the conducting surface,
which are themselves affected by the shapes of the molecules, leading
to even more complicated behavior and illustrating the need to treat
both atomistic and long-range many-body EM effects at these
scales~\footnote{Note that in principle, the response of the gold
  plate at nanometric separations should also be modified to account
  for atomic-scale effects, e.g. charge smearing, either
  phenomenologically by introducing a macroscopic nonlocal
  susceptibility (as is typically
  done~\cite{ChapuisPRB2008,SingerJQSRT2015,SchmidtJPB2018}) or
  atomistically within the proposed framework, but we leave this as
  subject of future work.}.

\emph{Concluding remarks.---} We have demonstrated a mesoscopic
approach to computing nonequilibrium thermal emission and radiative
heat transfer among molecules in the presence of planar metallic
surfaces, accounting for molecular nonlocal response as well as
many-body and multiple scattering effects to all orders. Our approach
demonstrates significant deviations in the emitted and transferred
power from commonly-used local/dipolar approximations, particularly in
the near-field and even at relatively large separations of
$10~\mathrm{nm}$, where the modification to long-range EM interactions
due to the proximate surface as well as the nonlocal molecular
susceptibility both matter; moreover, it has the virtue of being
derived from ab-initio DFT methods, in contrast to phenomenological
treatments of nonlocal
response~\cite{ChapuisPRB2008,SingerJQSRT2015,SchmidtJPB2018} which
give heat transfer results that depend strongly on the nonlocal
material model and boundary conditions considered. Our approach can be
generalized to consider extended materials like graphene or organic
crystals, which might enable accurate computations of radiative
contributions to bulk thermal conductivity beyond the Fourier regime,
as well as actively studied experimental situations, including
molecular materials in the vicinity of heated atomic force microscopy
tips which have applications to molecular junctions~\cite{CuiJCP2017,
  CuiSCIENCE2017}.

This work was supported by the National Science Foundation under
Grants No. DMR-1454836, DMR 1420541, and DGE 1148900, and the
Luxembourg National Research within the FNR-CORE program
(No. FNR-11360857). PSV thanks Chinmay Khandekar, Weiliang Jin, Sean
Molesky, and Theerachai Chanyaswad for the helpful discussions.





\section*{Appendix}


In what follows, we provide a compact derivation of \eqref{Phimn} in
the main text. Heat emission or transfer arises from the radiation of
fluctuating polarization sources $\ket{\vec{P}^{(0)}}$ within all
bodies, described by Maxwell's equations in integral form,
$\ket{\vec{E}} = \mathbb{G}_{0} \ket{\vec{P}}$,
$\ket{\vec{P}} = \ket{\vec{P}^{(0)}} + \left(\mathbb{V}_{\mathrm{env}}
  + \sum_{k = 1}^{N_{\mathrm{mol}}} \mathbb{V}_{\mathrm{k}} \right)
\ket{\vec{E}}$,
in terms of the vacuum, homogeneous Green's function
$\mathbb{G}_{0} (\omega, \vec{x}, \vec{x}') = \left(\nabla \otimes
  \nabla + \frac{\omega^{2}}{c^{2}} \mathbb{I}\right)
\frac{1}{4\pi|\vec{x}-\vec{x}'|} e^{\im\omega |\vec{x} -
  \vec{x}'|/c}$. Because we only consider incident fields arising out
of fluctuations within molecular bodies, the macroscopic degrees of
freedom $\mathbb{V}_{\mathrm{env}}$ can be traced out in favor of a
modified EM response or Green's function
$\mathbb{G}_{\mathrm{env}} = (\mathbb{G}_{0}^{-1} -
\mathbb{V}_{\mathrm{env}})^{-1}$ solving Maxwell's equations,
\begin{equation}
  \left[\nabla\times\nabla\times - \frac{\omega^{2}}{c^{2}} \left(\mathbb{I} + \mathbb{V}_{\mathrm{env}}\right)\right]\mathbb{G}_{\mathrm{env}} = \frac{\omega^{2}}{c^{2}} \mathbb{I}
\end{equation}
leading to modified field equations,
$\ket{\vec{E}} = \mathbb{G}_{\mathrm{env}} \ket{\vec{P}}$ and
$\ket{\vec{P}} = \ket{\vec{P}^{(0)}} + \mathbb{V}_{\mathrm{mol}}
\ket{\vec{E}}$, which depend only on the cumulative molecular
susceptibility
$\mathbb{V}_{\mathrm{mol}} = \sum_{k = 1}^{N_{\mathrm{mol}}}
\mathbb{V}_{k}$ and $\mathbb{G}_{\mathrm{env}}$. This allows the net
fields,
\begin{align}
  \ket{\vec{E}} &= \mathbb{G}_{\mathrm{env}} \mathbb{T}
  \mathbb{V}_{\mathrm{mol}}^{-1} \ket{\vec{P}^{(0)}} \\  
  \ket{\vec{P}} &= \mathbb{T} \mathbb{V}_{\mathrm{mol}}^{-1}
  \ket{\vec{P}^{(0)}},
\end{align}
to be expressed in terms of the scattering transition operator,
\begin{equation}
  \mathbb{T}^{-1} = \mathbb{V}_{\mathrm{mol}}^{-1} -
  \mathbb{G}_{\mathrm{env}},
\end{equation}
which describes EM scattering (to all orders) by the collection of
molecules in the presence of all
bodies~~\cite{RahiPRD2009,KrugerPRB2012}.


At this point, we take the (spatially uncorrelated) sources
$\ket{\vec{P}^{(0)}}$ to originate from a given molecule $m$, and
consider energy transfer to molecule $n$, such that
$\ket{\vec{P}^{(0)}} = \mathbb{P}_{m} \ket{\vec{P}^{(0)}_{m}}$, where
$\mathbb{P}_{m}$ projects onto the subspace of degrees of freedom for
molecule $m$. By Poynting's theorem, the energy transfer from $m$ to
$n$ is equal to the work done by the field,
\begin{equation}
  \Phi_{n}^{(m)} = -\frac{1}{2} \Re\left(\bracket{\mathbb{P}_{n}
    \vec{E}|\mathbb{P}_{n} \vec{J}}\right),
\end{equation}
where $\ket{\vec{J}} = -\im\omega\ket{\vec{P}}$ and
$\bracket{\vec{A}|\vec{B}} = \int \mathrm{d}^{3} x~\vec{A}^{\star}
(\vec{x}) \cdot \vec{B} (\vec{x})$. Using the above formulas for the
net EM and polarization fields, and exploiting the Hermiticity and
idempotence of projection operators under this inner product, allows
rewriting $\Phi_{n}^{(m)} = \frac{\omega}{4i}
\left(\bracket{\vec{E}|\mathbb{P}_{n} \vec{P}} -
\bracket{\mathbb{P}_{n} \vec{P}|\vec{E}}\right)$, which can be readily
expressed in terms of the scattering operators as,
\begin{equation}
\Phi_{n}^{(m)} = -\frac{\omega}{2}
\bracket{\vec{P}^{(0)}_{m}|\mathbb{P}_{m}
  \mathbb{V}_{\mathrm{mol}}^{-1 \star} \mathbb{T}^{\star}
  \asym\left(\mathbb{G}_{\mathrm{env}} \mathbb{P}_{n}\right)
  \mathbb{T} \mathbb{V}_{\mathrm{mol}}^{-1} \mathbb{P}_{m}
  \vec{P}^{(0)}_{m}}
\end{equation}
where we have exploited the symmetry (not Hermiticity) of
$\mathbb{V}_{\mathrm{mol}}$ and $\mathbb{T}$ under the conjugated
inner product. Here, ${}^{\star}$ implies complex conjugation,
${}^{\dagger}$ implies conjugate transposition (adjoint),
$\Im(\mathbb{A}) = \frac{\mathbb{A} - \mathbb{A}^{\star}}{2i}$, and
$\asym(\mathbb{A}) = \frac{\mathbb{A} - \mathbb{A}^{\dagger}}{2i}$.

Finally, we take the sources $\ket{\vec{P}^{(0)}_{m}}$ to arise from
thermal fluctuations. The fluctuation--dissipation theorem states that
for any polarizable body, thermal current fluctuations are related to
the dissipative part of the susceptibility via~\cite{Novotny2006}, 
\begin{equation}
  \bracket{\ket{\vec{P}^{(0)}_{m}} \bra{\vec{P}^{(0)}_{m}}} =
  \frac{1}{\pi\omega} \Im(\mathbb{V}_{m}),
\end{equation}
where the dependence on the Planck function
$\Theta(\omega, T) =
\hbar\omega/\left(\mathrm{exp}\left(\frac{\hbar\omega}{k_{\mathrm{B}}
      T}\right) - 1\right)$ (ignoring the zero-point term) has been
suppressed for now.  Using the identity,
$\mathbb{V}_{\mathrm{mol}}^{-1} \mathbb{P}_{m} \Im(\mathbb{V}_{m})
\mathbb{P}_{m} \mathbb{V}_{\mathrm{mol}}^{-1 \star} = \mathbb{P}_{m}
\mathbb{V}_{m}^{-1} \Im(\mathbb{V}_{m}) \mathbb{V}_{m}^{-1\star}
\mathbb{P}_{m} = \mathbb{P}_{m} \Im(\mathbb{V}_{m}^{-1 \star})
\mathbb{P}_{m}$, it follows that the energy transfer can be written as
the trace expression given in \eqref{Phimn} of the main text. Note
that a similar expression may be obtained for the energy transfer
between a molecule and a macroscopic body, the evaluation of which
requires application of more complicated scattering
techniques~\cite{PolimeridisPRB2015, RodriguezPRB2013}, the subject of
future work.

\bibliography{molheatpaper2}

\begin{thebibliography}{75}%
\makeatletter
\providecommand \@ifxundefined [1]{%
 \@ifx{#1\undefined}
}%
\providecommand \@ifnum [1]{%
 \ifnum #1\expandafter \@firstoftwo
 \else \expandafter \@secondoftwo
 \fi
}%
\providecommand \@ifx [1]{%
 \ifx #1\expandafter \@firstoftwo
 \else \expandafter \@secondoftwo
 \fi
}%
\providecommand \natexlab [1]{#1}%
\providecommand \enquote  [1]{``#1''}%
\providecommand \bibnamefont  [1]{#1}%
\providecommand \bibfnamefont [1]{#1}%
\providecommand \citenamefont [1]{#1}%
\providecommand \href@noop [0]{\@secondoftwo}%
\providecommand \href [0]{\begingroup \@sanitize@url \@href}%
\providecommand \@href[1]{\@@startlink{#1}\@@href}%
\providecommand \@@href[1]{\endgroup#1\@@endlink}%
\providecommand \@sanitize@url [0]{\catcode `\\12\catcode `\$12\catcode
  `\&12\catcode `\#12\catcode `\^12\catcode `\_12\catcode `\%12\relax}%
\providecommand \@@startlink[1]{}%
\providecommand \@@endlink[0]{}%
\providecommand \url  [0]{\begingroup\@sanitize@url \@url }%
\providecommand \@url [1]{\endgroup\@href {#1}{\urlprefix }}%
\providecommand \urlprefix  [0]{URL }%
\providecommand \Eprint [0]{\href }%
\providecommand \doibase [0]{http://dx.doi.org/}%
\providecommand \selectlanguage [0]{\@gobble}%
\providecommand \bibinfo  [0]{\@secondoftwo}%
\providecommand \bibfield  [0]{\@secondoftwo}%
\providecommand \translation [1]{[#1]}%
\providecommand \BibitemOpen [0]{}%
\providecommand \bibitemStop [0]{}%
\providecommand \bibitemNoStop [0]{.\EOS\space}%
\providecommand \EOS [0]{\spacefactor3000\relax}%
\providecommand \BibitemShut  [1]{\csname bibitem#1\endcsname}%
\let\auto@bib@innerbib\@empty
\bibitem [{\citenamefont {Messina}\ and\ \citenamefont
  {Ben-Abdallah}(2013)}]{MessinaNATURE2013}%
  \BibitemOpen
  \bibfield  {author} {\bibinfo {author} {\bibfnamefont {R.}~\bibnamefont
  {Messina}}\ and\ \bibinfo {author} {\bibfnamefont {P.}~\bibnamefont
  {Ben-Abdallah}},\ }\href {http://dx.doi.org/10.1038/srep01383} {\ \textbf
  {\bibinfo {volume} {3}},\ \bibinfo {pages} {1383} (\bibinfo {year}
  {2013})}\BibitemShut {NoStop}%
\bibitem [{\citenamefont {Svetovoy}\ and\ \citenamefont
  {Palasantzas}(2014)}]{SvetovoyPRAPP2014}%
  \BibitemOpen
  \bibfield  {author} {\bibinfo {author} {\bibfnamefont {V.~B.}\ \bibnamefont
  {Svetovoy}}\ and\ \bibinfo {author} {\bibfnamefont {G.}~\bibnamefont
  {Palasantzas}},\ }\href {\doibase 10.1103/PhysRevApplied.2.034006} {\bibfield
   {journal} {\bibinfo  {journal} {Phys. Rev. Applied}\ }\textbf {\bibinfo
  {volume} {2}},\ \bibinfo {pages} {034006} (\bibinfo {year}
  {2014})}\BibitemShut {NoStop}%
\bibitem [{\citenamefont {Ilic}\ \emph {et~al.}(2012)\citenamefont {Ilic},
  \citenamefont {Jablan}, \citenamefont {Joannopoulos}, \citenamefont
  {Celanovic},\ and\ \citenamefont {Solja\v{c}i\'{c}}}]{IlicOE2012}%
  \BibitemOpen
  \bibfield  {author} {\bibinfo {author} {\bibfnamefont {O.}~\bibnamefont
  {Ilic}}, \bibinfo {author} {\bibfnamefont {M.}~\bibnamefont {Jablan}},
  \bibinfo {author} {\bibfnamefont {J.~D.}\ \bibnamefont {Joannopoulos}},
  \bibinfo {author} {\bibfnamefont {I.}~\bibnamefont {Celanovic}}, \ and\
  \bibinfo {author} {\bibfnamefont {M.}~\bibnamefont {Solja\v{c}i\'{c}}},\
  }\href {\doibase 10.1364/OE.20.00A366} {\bibfield  {journal} {\bibinfo
  {journal} {Opt. Express}\ }\textbf {\bibinfo {volume} {20}},\ \bibinfo
  {pages} {A366} (\bibinfo {year} {2012})}\BibitemShut {NoStop}%
\bibitem [{\citenamefont {Lenert}\ \emph {et~al.}(2014)\citenamefont {Lenert},
  \citenamefont {Bierman}, \citenamefont {Nam}, \citenamefont {Chan},
  \citenamefont {Celanovi{\'c}}, \citenamefont {Solja{\v{c}}i{\'c}},\ and\
  \citenamefont {Wang}}]{LenertNATURENANO2014}%
  \BibitemOpen
  \bibfield  {author} {\bibinfo {author} {\bibfnamefont {A.}~\bibnamefont
  {Lenert}}, \bibinfo {author} {\bibfnamefont {D.~M.}\ \bibnamefont {Bierman}},
  \bibinfo {author} {\bibfnamefont {Y.}~\bibnamefont {Nam}}, \bibinfo {author}
  {\bibfnamefont {W.~R.}\ \bibnamefont {Chan}}, \bibinfo {author}
  {\bibfnamefont {I.}~\bibnamefont {Celanovi{\'c}}}, \bibinfo {author}
  {\bibfnamefont {M.}~\bibnamefont {Solja{\v{c}}i{\'c}}}, \ and\ \bibinfo
  {author} {\bibfnamefont {E.~N.}\ \bibnamefont {Wang}},\ }\href@noop {}
  {\bibfield  {journal} {\bibinfo  {journal} {Nature nanotechnology}\ }\textbf
  {\bibinfo {volume} {9}},\ \bibinfo {pages} {126} (\bibinfo {year}
  {2014})}\BibitemShut {NoStop}%
\bibitem [{\citenamefont {Ruan}\ and\ \citenamefont
  {Kaviany}(2006)}]{RuanPRB2006}%
  \BibitemOpen
  \bibfield  {author} {\bibinfo {author} {\bibfnamefont {X.~L.}\ \bibnamefont
  {Ruan}}\ and\ \bibinfo {author} {\bibfnamefont {M.}~\bibnamefont {Kaviany}},\
  }\href {\doibase 10.1103/PhysRevB.73.155422} {\bibfield  {journal} {\bibinfo
  {journal} {Phys. Rev. B}\ }\textbf {\bibinfo {volume} {73}},\ \bibinfo
  {pages} {155422} (\bibinfo {year} {2006})}\BibitemShut {NoStop}%
\bibitem [{\citenamefont {Boriskina}\ \emph {et~al.}(2016)\citenamefont
  {Boriskina}, \citenamefont {Tong}, \citenamefont {Hsu}, \citenamefont {Liao},
  \citenamefont {Huang}, \citenamefont {Chiloyan},\ and\ \citenamefont
  {Chen}}]{BoriskinaNANOPHOT2016}%
  \BibitemOpen
  \bibfield  {author} {\bibinfo {author} {\bibfnamefont {S.~V.}\ \bibnamefont
  {Boriskina}}, \bibinfo {author} {\bibfnamefont {J.~K.}\ \bibnamefont {Tong}},
  \bibinfo {author} {\bibfnamefont {W.-C.}\ \bibnamefont {Hsu}}, \bibinfo
  {author} {\bibfnamefont {B.}~\bibnamefont {Liao}}, \bibinfo {author}
  {\bibfnamefont {Y.}~\bibnamefont {Huang}}, \bibinfo {author} {\bibfnamefont
  {V.}~\bibnamefont {Chiloyan}}, \ and\ \bibinfo {author} {\bibfnamefont
  {G.}~\bibnamefont {Chen}},\ }\href@noop {} {\bibfield  {journal} {\bibinfo
  {journal} {Nanophotonics}\ }\textbf {\bibinfo {volume} {5}},\ \bibinfo
  {pages} {134} (\bibinfo {year} {2016})}\BibitemShut {NoStop}%
\bibitem [{\citenamefont {Cui}\ \emph {et~al.}(2017{\natexlab{a}})\citenamefont
  {Cui}, \citenamefont {Miao}, \citenamefont {Jiang}, \citenamefont
  {Meyhofer},\ and\ \citenamefont {Reddy}}]{CuiJCP2017}%
  \BibitemOpen
  \bibfield  {author} {\bibinfo {author} {\bibfnamefont {L.}~\bibnamefont
  {Cui}}, \bibinfo {author} {\bibfnamefont {R.}~\bibnamefont {Miao}}, \bibinfo
  {author} {\bibfnamefont {C.}~\bibnamefont {Jiang}}, \bibinfo {author}
  {\bibfnamefont {E.}~\bibnamefont {Meyhofer}}, \ and\ \bibinfo {author}
  {\bibfnamefont {P.}~\bibnamefont {Reddy}},\ }\href@noop {} {\bibfield
  {journal} {\bibinfo  {journal} {The Journal of Chemical Physics}\ }\textbf
  {\bibinfo {volume} {146}},\ \bibinfo {pages} {092201} (\bibinfo {year}
  {2017}{\natexlab{a}})}\BibitemShut {NoStop}%
\bibitem [{\citenamefont {Cui}\ \emph {et~al.}(2017{\natexlab{b}})\citenamefont
  {Cui}, \citenamefont {Jeong}, \citenamefont {Hur}, \citenamefont {Matt},
  \citenamefont {Kl{\"o}ckner}, \citenamefont {Pauly}, \citenamefont {Nielaba},
  \citenamefont {Cuevas}, \citenamefont {Meyhofer},\ and\ \citenamefont
  {Reddy}}]{CuiSCIENCE2017}%
  \BibitemOpen
  \bibfield  {author} {\bibinfo {author} {\bibfnamefont {L.}~\bibnamefont
  {Cui}}, \bibinfo {author} {\bibfnamefont {W.}~\bibnamefont {Jeong}}, \bibinfo
  {author} {\bibfnamefont {S.}~\bibnamefont {Hur}}, \bibinfo {author}
  {\bibfnamefont {M.}~\bibnamefont {Matt}}, \bibinfo {author} {\bibfnamefont
  {J.~C.}\ \bibnamefont {Kl{\"o}ckner}}, \bibinfo {author} {\bibfnamefont
  {F.}~\bibnamefont {Pauly}}, \bibinfo {author} {\bibfnamefont
  {P.}~\bibnamefont {Nielaba}}, \bibinfo {author} {\bibfnamefont {J.~C.}\
  \bibnamefont {Cuevas}}, \bibinfo {author} {\bibfnamefont {E.}~\bibnamefont
  {Meyhofer}}, \ and\ \bibinfo {author} {\bibfnamefont {P.}~\bibnamefont
  {Reddy}},\ }\href {\doibase 10.1126/science.aam6622} {\  (\bibinfo {year}
  {2017}{\natexlab{b}}),\ 10.1126/science.aam6622}\BibitemShut {NoStop}%
\bibitem [{\citenamefont {Cui}\ \emph {et~al.}(2015)\citenamefont {Cui},
  \citenamefont {Feng},\ and\ \citenamefont {Zhang}}]{CuiJPCA2015}%
  \BibitemOpen
  \bibfield  {author} {\bibinfo {author} {\bibfnamefont {L.}~\bibnamefont
  {Cui}}, \bibinfo {author} {\bibfnamefont {Y.}~\bibnamefont {Feng}}, \ and\
  \bibinfo {author} {\bibfnamefont {X.}~\bibnamefont {Zhang}},\ }\href
  {\doibase 10.1021/acs.jpca.5b07995} {\bibfield  {journal} {\bibinfo
  {journal} {The Journal of Physical Chemistry A}\ }\textbf {\bibinfo {volume}
  {119}},\ \bibinfo {pages} {11226} (\bibinfo {year} {2015})}\BibitemShut
  {NoStop}%
\bibitem [{\citenamefont {Pendry}\ \emph {et~al.}(2016)\citenamefont {Pendry},
  \citenamefont {Sasihithlu},\ and\ \citenamefont {Craster}}]{PendryPRB2016}%
  \BibitemOpen
  \bibfield  {author} {\bibinfo {author} {\bibfnamefont {J.~B.}\ \bibnamefont
  {Pendry}}, \bibinfo {author} {\bibfnamefont {K.}~\bibnamefont {Sasihithlu}},
  \ and\ \bibinfo {author} {\bibfnamefont {R.~V.}\ \bibnamefont {Craster}},\
  }\href {\doibase 10.1103/PhysRevB.94.075414} {\bibfield  {journal} {\bibinfo
  {journal} {Phys. Rev. B}\ }\textbf {\bibinfo {volume} {94}},\ \bibinfo
  {pages} {075414} (\bibinfo {year} {2016})}\BibitemShut {NoStop}%
\bibitem [{\citenamefont {Edalatpour}\ and\ \citenamefont
  {Francoeur}(2016)}]{EdalatpourPRB2016}%
  \BibitemOpen
  \bibfield  {author} {\bibinfo {author} {\bibfnamefont {S.}~\bibnamefont
  {Edalatpour}}\ and\ \bibinfo {author} {\bibfnamefont {M.}~\bibnamefont
  {Francoeur}},\ }\href {\doibase 10.1103/PhysRevB.94.045406} {\bibfield
  {journal} {\bibinfo  {journal} {Phys. Rev. B}\ }\textbf {\bibinfo {volume}
  {94}},\ \bibinfo {pages} {045406} (\bibinfo {year} {2016})}\BibitemShut
  {NoStop}%
\bibitem [{\citenamefont {Chiloyan}\ \emph {et~al.}(2015)\citenamefont
  {Chiloyan}, \citenamefont {Garg}, \citenamefont {Esfarjani},\ and\
  \citenamefont {Chen}}]{ChiloyanNATURE2015}%
  \BibitemOpen
  \bibfield  {author} {\bibinfo {author} {\bibfnamefont {V.}~\bibnamefont
  {Chiloyan}}, \bibinfo {author} {\bibfnamefont {J.}~\bibnamefont {Garg}},
  \bibinfo {author} {\bibfnamefont {K.}~\bibnamefont {Esfarjani}}, \ and\
  \bibinfo {author} {\bibfnamefont {G.}~\bibnamefont {Chen}},\ }\href@noop {}
  {\bibfield  {journal} {\bibinfo  {journal} {Nature communications}\ }\textbf
  {\bibinfo {volume} {6}},\ \bibinfo {pages} {6755} (\bibinfo {year}
  {2015})}\BibitemShut {NoStop}%
\bibitem [{\citenamefont {Kloppstech}\ \emph {et~al.}(2017)\citenamefont
  {Kloppstech}, \citenamefont {K{\"o}nne}, \citenamefont {Biehs}, \citenamefont
  {Rodriguez}, \citenamefont {Worbes}, \citenamefont {Hellmann},\ and\
  \citenamefont {Kittel}}]{KloppstechNATURE2017}%
  \BibitemOpen
  \bibfield  {author} {\bibinfo {author} {\bibfnamefont {K.}~\bibnamefont
  {Kloppstech}}, \bibinfo {author} {\bibfnamefont {N.}~\bibnamefont
  {K{\"o}nne}}, \bibinfo {author} {\bibfnamefont {S.-A.}\ \bibnamefont
  {Biehs}}, \bibinfo {author} {\bibfnamefont {A.~W.}\ \bibnamefont
  {Rodriguez}}, \bibinfo {author} {\bibfnamefont {L.}~\bibnamefont {Worbes}},
  \bibinfo {author} {\bibfnamefont {D.}~\bibnamefont {Hellmann}}, \ and\
  \bibinfo {author} {\bibfnamefont {A.}~\bibnamefont {Kittel}},\ }\href@noop {}
  {\bibfield  {journal} {\bibinfo  {journal} {Nature Communications}\ }\textbf
  {\bibinfo {volume} {8}} (\bibinfo {year} {2017})}\BibitemShut {NoStop}%
\bibitem [{\citenamefont {Cui}\ \emph {et~al.}(2017{\natexlab{c}})\citenamefont
  {Cui}, \citenamefont {Jeong}, \citenamefont {Fern{\'a}ndez-Hurtado},
  \citenamefont {Feist}, \citenamefont {Garc{\'\i}a-Vidal}, \citenamefont
  {Cuevas}, \citenamefont {Meyhofer},\ and\ \citenamefont
  {Reddy}}]{CuiNATURE2017}%
  \BibitemOpen
  \bibfield  {author} {\bibinfo {author} {\bibfnamefont {L.}~\bibnamefont
  {Cui}}, \bibinfo {author} {\bibfnamefont {W.}~\bibnamefont {Jeong}}, \bibinfo
  {author} {\bibfnamefont {V.}~\bibnamefont {Fern{\'a}ndez-Hurtado}}, \bibinfo
  {author} {\bibfnamefont {J.}~\bibnamefont {Feist}}, \bibinfo {author}
  {\bibfnamefont {F.~J.}\ \bibnamefont {Garc{\'\i}a-Vidal}}, \bibinfo {author}
  {\bibfnamefont {J.~C.}\ \bibnamefont {Cuevas}}, \bibinfo {author}
  {\bibfnamefont {E.}~\bibnamefont {Meyhofer}}, \ and\ \bibinfo {author}
  {\bibfnamefont {P.}~\bibnamefont {Reddy}},\ }\href@noop {} {\bibfield
  {journal} {\bibinfo  {journal} {Nature Communications}\ }\textbf {\bibinfo
  {volume} {8}} (\bibinfo {year} {2017}{\natexlab{c}})}\BibitemShut {NoStop}%
\bibitem [{\citenamefont {St-Gelais}\ \emph {et~al.}(2014)\citenamefont
  {St-Gelais}, \citenamefont {Guha}, \citenamefont {Zhu}, \citenamefont {Fan},\
  and\ \citenamefont {Lipson}}]{StGelaisNANOLETT2014}%
  \BibitemOpen
  \bibfield  {author} {\bibinfo {author} {\bibfnamefont {R.}~\bibnamefont
  {St-Gelais}}, \bibinfo {author} {\bibfnamefont {B.}~\bibnamefont {Guha}},
  \bibinfo {author} {\bibfnamefont {L.}~\bibnamefont {Zhu}}, \bibinfo {author}
  {\bibfnamefont {S.}~\bibnamefont {Fan}}, \ and\ \bibinfo {author}
  {\bibfnamefont {M.}~\bibnamefont {Lipson}},\ }\href {\doibase
  10.1021/nl503236k} {\bibfield  {journal} {\bibinfo  {journal} {Nano Letters}\
  }\textbf {\bibinfo {volume} {14}},\ \bibinfo {pages} {6971} (\bibinfo {year}
  {2014})},\ \Eprint {http://arxiv.org/abs/http://dx.doi.org/10.1021/nl503236k}
  {http://dx.doi.org/10.1021/nl503236k} \BibitemShut {NoStop}%
\bibitem [{\citenamefont {Tian}\ \emph {et~al.}(2014)\citenamefont {Tian},
  \citenamefont {Esfarjani},\ and\ \citenamefont {Chen}}]{TianPRB2014}%
  \BibitemOpen
  \bibfield  {author} {\bibinfo {author} {\bibfnamefont {Z.}~\bibnamefont
  {Tian}}, \bibinfo {author} {\bibfnamefont {K.}~\bibnamefont {Esfarjani}}, \
  and\ \bibinfo {author} {\bibfnamefont {G.}~\bibnamefont {Chen}},\ }\href
  {\doibase 10.1103/PhysRevB.89.235307} {\bibfield  {journal} {\bibinfo
  {journal} {Phys. Rev. B}\ }\textbf {\bibinfo {volume} {89}},\ \bibinfo
  {pages} {235307} (\bibinfo {year} {2014})}\BibitemShut {NoStop}%
\bibitem [{\citenamefont {Tian}\ \emph {et~al.}(2012)\citenamefont {Tian},
  \citenamefont {Esfarjani},\ and\ \citenamefont {Chen}}]{TianPRB2012}%
  \BibitemOpen
  \bibfield  {author} {\bibinfo {author} {\bibfnamefont {Z.}~\bibnamefont
  {Tian}}, \bibinfo {author} {\bibfnamefont {K.}~\bibnamefont {Esfarjani}}, \
  and\ \bibinfo {author} {\bibfnamefont {G.}~\bibnamefont {Chen}},\ }\href
  {\doibase 10.1103/PhysRevB.86.235304} {\bibfield  {journal} {\bibinfo
  {journal} {Phys. Rev. B}\ }\textbf {\bibinfo {volume} {86}},\ \bibinfo
  {pages} {235304} (\bibinfo {year} {2012})}\BibitemShut {NoStop}%
\bibitem [{\citenamefont {Rodriguez}\ \emph {et~al.}(2013)\citenamefont
  {Rodriguez}, \citenamefont {Reid},\ and\ \citenamefont
  {Johnson}}]{RodriguezPRB2013}%
  \BibitemOpen
  \bibfield  {author} {\bibinfo {author} {\bibfnamefont {A.~W.}\ \bibnamefont
  {Rodriguez}}, \bibinfo {author} {\bibfnamefont {M.~T.~H.}\ \bibnamefont
  {Reid}}, \ and\ \bibinfo {author} {\bibfnamefont {S.~G.}\ \bibnamefont
  {Johnson}},\ }\href {\doibase 10.1103/PhysRevB.88.054305} {\bibfield
  {journal} {\bibinfo  {journal} {Phys. Rev. B}\ }\textbf {\bibinfo {volume}
  {88}},\ \bibinfo {pages} {054305} (\bibinfo {year} {2013})}\BibitemShut
  {NoStop}%
\bibitem [{\citenamefont {Polimeridis}\ \emph {et~al.}(2015)\citenamefont
  {Polimeridis}, \citenamefont {Reid}, \citenamefont {Jin}, \citenamefont
  {Johnson}, \citenamefont {White},\ and\ \citenamefont
  {Rodriguez}}]{PolimeridisPRB2015}%
  \BibitemOpen
  \bibfield  {author} {\bibinfo {author} {\bibfnamefont {A.~G.}\ \bibnamefont
  {Polimeridis}}, \bibinfo {author} {\bibfnamefont {M.~T.~H.}\ \bibnamefont
  {Reid}}, \bibinfo {author} {\bibfnamefont {W.}~\bibnamefont {Jin}}, \bibinfo
  {author} {\bibfnamefont {S.~G.}\ \bibnamefont {Johnson}}, \bibinfo {author}
  {\bibfnamefont {J.~K.}\ \bibnamefont {White}}, \ and\ \bibinfo {author}
  {\bibfnamefont {A.~W.}\ \bibnamefont {Rodriguez}},\ }\href {\doibase
  10.1103/PhysRevB.92.134202} {\bibfield  {journal} {\bibinfo  {journal} {Phys.
  Rev. B}\ }\textbf {\bibinfo {volume} {92}},\ \bibinfo {pages} {134202}
  (\bibinfo {year} {2015})}\BibitemShut {NoStop}%
\bibitem [{\citenamefont {Hajian}\ \emph {et~al.}(2017)\citenamefont {Hajian},
  \citenamefont {Ghobadi}, \citenamefont {Dereshgi}, \citenamefont {Butun},\
  and\ \citenamefont {Ozbay}}]{HajianJOSAB2017}%
  \BibitemOpen
  \bibfield  {author} {\bibinfo {author} {\bibfnamefont {H.}~\bibnamefont
  {Hajian}}, \bibinfo {author} {\bibfnamefont {A.}~\bibnamefont {Ghobadi}},
  \bibinfo {author} {\bibfnamefont {S.~A.}\ \bibnamefont {Dereshgi}}, \bibinfo
  {author} {\bibfnamefont {B.}~\bibnamefont {Butun}}, \ and\ \bibinfo {author}
  {\bibfnamefont {E.}~\bibnamefont {Ozbay}},\ }\href {\doibase
  10.1364/JOSAB.34.000D29} {\bibfield  {journal} {\bibinfo  {journal} {J. Opt.
  Soc. Am. B}\ }\textbf {\bibinfo {volume} {34}},\ \bibinfo {pages} {D29}
  (\bibinfo {year} {2017})}\BibitemShut {NoStop}%
\bibitem [{\citenamefont {Shen}\ \emph {et~al.}(2009)\citenamefont {Shen},
  \citenamefont {Narayanaswamy},\ and\ \citenamefont
  {Chen}}]{ShenNANOLETT2009}%
  \BibitemOpen
  \bibfield  {author} {\bibinfo {author} {\bibfnamefont {S.}~\bibnamefont
  {Shen}}, \bibinfo {author} {\bibfnamefont {A.}~\bibnamefont {Narayanaswamy}},
  \ and\ \bibinfo {author} {\bibfnamefont {G.}~\bibnamefont {Chen}},\ }\href
  {\doibase 10.1021/nl901208v} {\bibfield  {journal} {\bibinfo  {journal} {Nano
  Letters}\ }\textbf {\bibinfo {volume} {9}},\ \bibinfo {pages} {2909}
  (\bibinfo {year} {2009})},\ \Eprint
  {http://arxiv.org/abs/http://dx.doi.org/10.1021/nl901208v}
  {http://dx.doi.org/10.1021/nl901208v} \BibitemShut {NoStop}%
\bibitem [{\citenamefont {Garc{\'\i}a~de
  Abajo}(2013)}]{GarciaDeAbajoACSNANO2013}%
  \BibitemOpen
  \bibfield  {author} {\bibinfo {author} {\bibfnamefont {F.~J.}\ \bibnamefont
  {Garc{\'\i}a~de Abajo}},\ }\href {\doibase 10.1021/nn405367e} {\bibfield
  {journal} {\bibinfo  {journal} {ACS Nano}\ }\textbf {\bibinfo {volume} {7}},\
  \bibinfo {pages} {11409} (\bibinfo {year} {2013})},\ \Eprint
  {http://arxiv.org/abs/http://dx.doi.org/10.1021/nn405367e}
  {http://dx.doi.org/10.1021/nn405367e} \BibitemShut {NoStop}%
\bibitem [{\citenamefont {Basov}\ \emph {et~al.}(2016)\citenamefont {Basov},
  \citenamefont {Fogler},\ and\ \citenamefont {Garc{\'\i}a~de
  Abajo}}]{BasovSCIENCE2016}%
  \BibitemOpen
  \bibfield  {author} {\bibinfo {author} {\bibfnamefont {D.~N.}\ \bibnamefont
  {Basov}}, \bibinfo {author} {\bibfnamefont {M.~M.}\ \bibnamefont {Fogler}}, \
  and\ \bibinfo {author} {\bibfnamefont {F.~J.}\ \bibnamefont {Garc{\'\i}a~de
  Abajo}},\ }\href {\doibase 10.1126/science.aag1992} {\ \textbf {\bibinfo
  {volume} {354}} (\bibinfo {year} {2016}),\
  10.1126/science.aag1992}\BibitemShut {NoStop}%
\bibitem [{\citenamefont {Luo}\ and\ \citenamefont {Chen}(2013)}]{LuoPCCP2013}%
  \BibitemOpen
  \bibfield  {author} {\bibinfo {author} {\bibfnamefont {T.}~\bibnamefont
  {Luo}}\ and\ \bibinfo {author} {\bibfnamefont {G.}~\bibnamefont {Chen}},\
  }\href {\doibase 10.1039/C2CP43771F} {\bibfield  {journal} {\bibinfo
  {journal} {Phys. Chem. Chem. Phys.}\ }\textbf {\bibinfo {volume} {15}},\
  \bibinfo {pages} {3389} (\bibinfo {year} {2013})}\BibitemShut {NoStop}%
\bibitem [{\citenamefont {Carey}\ \emph {et~al.}(2008)\citenamefont {Carey},
  \citenamefont {Chen}, \citenamefont {Grigoropoulos}, \citenamefont
  {Kaviany},\ and\ \citenamefont {Majumdar}}]{CareyNMTE2008}%
  \BibitemOpen
  \bibfield  {author} {\bibinfo {author} {\bibfnamefont {V.~P.}\ \bibnamefont
  {Carey}}, \bibinfo {author} {\bibfnamefont {G.}~\bibnamefont {Chen}},
  \bibinfo {author} {\bibfnamefont {C.}~\bibnamefont {Grigoropoulos}}, \bibinfo
  {author} {\bibfnamefont {M.}~\bibnamefont {Kaviany}}, \ and\ \bibinfo
  {author} {\bibfnamefont {A.}~\bibnamefont {Majumdar}},\ }\href {\doibase
  10.1080/15567260801917520} {\bibfield  {journal} {\bibinfo  {journal}
  {Nanoscale and Microscale Thermophysical Engineering}\ }\textbf {\bibinfo
  {volume} {12}},\ \bibinfo {pages} {1} (\bibinfo {year} {2008})}\BibitemShut
  {NoStop}%
\bibitem [{\citenamefont {Boriskina}\ \emph {et~al.}(2013)\citenamefont
  {Boriskina}, \citenamefont {Ghasemi},\ and\ \citenamefont
  {Chen}}]{BoriskinaMT2013}%
  \BibitemOpen
  \bibfield  {author} {\bibinfo {author} {\bibfnamefont {S.~V.}\ \bibnamefont
  {Boriskina}}, \bibinfo {author} {\bibfnamefont {H.}~\bibnamefont {Ghasemi}},
  \ and\ \bibinfo {author} {\bibfnamefont {G.}~\bibnamefont {Chen}},\ }\href
  {\doibase https://doi.org/10.1016/j.mattod.2013.09.003} {\bibfield  {journal}
  {\bibinfo  {journal} {Materials Today}\ }\textbf {\bibinfo {volume} {16}},\
  \bibinfo {pages} {375} (\bibinfo {year} {2013})}\BibitemShut {NoStop}%
\bibitem [{\citenamefont {Cahill}\ \emph {et~al.}(2014)\citenamefont {Cahill},
  \citenamefont {Braun}, \citenamefont {Chen}, \citenamefont {Clarke},
  \citenamefont {Fan}, \citenamefont {Goodson}, \citenamefont {Keblinski},
  \citenamefont {King}, \citenamefont {Mahan}, \citenamefont {Majumdar},
  \citenamefont {Maris}, \citenamefont {Phillpot}, \citenamefont {Pop},\ and\
  \citenamefont {Shi}}]{CahillAPR2014}%
  \BibitemOpen
  \bibfield  {author} {\bibinfo {author} {\bibfnamefont {D.~G.}\ \bibnamefont
  {Cahill}}, \bibinfo {author} {\bibfnamefont {P.~V.}\ \bibnamefont {Braun}},
  \bibinfo {author} {\bibfnamefont {G.}~\bibnamefont {Chen}}, \bibinfo {author}
  {\bibfnamefont {D.~R.}\ \bibnamefont {Clarke}}, \bibinfo {author}
  {\bibfnamefont {S.}~\bibnamefont {Fan}}, \bibinfo {author} {\bibfnamefont
  {K.~E.}\ \bibnamefont {Goodson}}, \bibinfo {author} {\bibfnamefont
  {P.}~\bibnamefont {Keblinski}}, \bibinfo {author} {\bibfnamefont {W.~P.}\
  \bibnamefont {King}}, \bibinfo {author} {\bibfnamefont {G.~D.}\ \bibnamefont
  {Mahan}}, \bibinfo {author} {\bibfnamefont {A.}~\bibnamefont {Majumdar}},
  \bibinfo {author} {\bibfnamefont {H.~J.}\ \bibnamefont {Maris}}, \bibinfo
  {author} {\bibfnamefont {S.~R.}\ \bibnamefont {Phillpot}}, \bibinfo {author}
  {\bibfnamefont {E.}~\bibnamefont {Pop}}, \ and\ \bibinfo {author}
  {\bibfnamefont {L.}~\bibnamefont {Shi}},\ }\href {\doibase 10.1063/1.4832615}
  {\bibfield  {journal} {\bibinfo  {journal} {Applied Physics Reviews}\
  }\textbf {\bibinfo {volume} {1}},\ \bibinfo {pages} {011305} (\bibinfo {year}
  {2014})},\ \Eprint {http://arxiv.org/abs/http://dx.doi.org/10.1063/1.4832615}
  {http://dx.doi.org/10.1063/1.4832615} \BibitemShut {NoStop}%
\bibitem [{\citenamefont {Dhar}\ and\ \citenamefont {Roy}(2006)}]{DharJSP2006}%
  \BibitemOpen
  \bibfield  {author} {\bibinfo {author} {\bibfnamefont {A.}~\bibnamefont
  {Dhar}}\ and\ \bibinfo {author} {\bibfnamefont {D.}~\bibnamefont {Roy}},\
  }\href {\doibase 10.1007/s10955-006-9235-3} {\bibfield  {journal} {\bibinfo
  {journal} {Journal of Statistical Physics}\ }\textbf {\bibinfo {volume}
  {125}},\ \bibinfo {pages} {801} (\bibinfo {year} {2006})}\BibitemShut
  {NoStop}%
\bibitem [{\citenamefont {Mingo}\ and\ \citenamefont
  {Yang}(2003)}]{MingoPRB2003}%
  \BibitemOpen
  \bibfield  {author} {\bibinfo {author} {\bibfnamefont {N.}~\bibnamefont
  {Mingo}}\ and\ \bibinfo {author} {\bibfnamefont {L.}~\bibnamefont {Yang}},\
  }\href {\doibase 10.1103/PhysRevB.68.245406} {\bibfield  {journal} {\bibinfo
  {journal} {Phys. Rev. B}\ }\textbf {\bibinfo {volume} {68}},\ \bibinfo
  {pages} {245406} (\bibinfo {year} {2003})}\BibitemShut {NoStop}%
\bibitem [{\citenamefont {Tkatchenko}\ \emph {et~al.}(2013)\citenamefont
  {Tkatchenko}, \citenamefont {Ambrosetti},\ and\ \citenamefont {{DiStasio
  Jr.}}}]{TkatchenkoJCP2013}%
  \BibitemOpen
  \bibfield  {author} {\bibinfo {author} {\bibfnamefont {A.}~\bibnamefont
  {Tkatchenko}}, \bibinfo {author} {\bibfnamefont {A.}~\bibnamefont
  {Ambrosetti}}, \ and\ \bibinfo {author} {\bibfnamefont {R.~A.}\ \bibnamefont
  {{DiStasio Jr.}}},\ }\href@noop {} {\bibfield  {journal} {\bibinfo  {journal}
  {The Journal of Chemical Physics}\ }\textbf {\bibinfo {volume} {138}}
  (\bibinfo {year} {2013})}\BibitemShut {NoStop}%
\bibitem [{\citenamefont {Gobre}\ and\ \citenamefont
  {Tkatchenko}(2013)}]{GobreNCOMMS2013}%
  \BibitemOpen
  \bibfield  {author} {\bibinfo {author} {\bibfnamefont {V.~V.}\ \bibnamefont
  {Gobre}}\ and\ \bibinfo {author} {\bibfnamefont {A.}~\bibnamefont
  {Tkatchenko}},\ }\href {\doibase http://dx.doi.org/10.1038/ncomms3341}
  {\bibfield  {journal} {\bibinfo  {journal} {Nature Communications}\ }\textbf
  {\bibinfo {volume} {4}} (\bibinfo {year} {2013}),\
  http://dx.doi.org/10.1038/ncomms3341}\BibitemShut {NoStop}%
\bibitem [{\citenamefont {{DiStasio Jr.}}\ \emph {et~al.}(2014)\citenamefont
  {{DiStasio Jr.}}, \citenamefont {Gobre},\ and\ \citenamefont
  {Tkatchenko}}]{DiStasioJPCM2014}%
  \BibitemOpen
  \bibfield  {author} {\bibinfo {author} {\bibfnamefont {R.~A.}\ \bibnamefont
  {{DiStasio Jr.}}}, \bibinfo {author} {\bibfnamefont {V.~V.}\ \bibnamefont
  {Gobre}}, \ and\ \bibinfo {author} {\bibfnamefont {A.}~\bibnamefont
  {Tkatchenko}},\ }\href {http://stacks.iop.org/0953-8984/26/i=21/a=213202}
  {\bibfield  {journal} {\bibinfo  {journal} {Journal of Physics: Condensed
  Matter}\ }\textbf {\bibinfo {volume} {26}},\ \bibinfo {pages} {213202}
  (\bibinfo {year} {2014})}\BibitemShut {NoStop}%
\bibitem [{\citenamefont {Tkatchenko}(2015)}]{TkatchenkoADFM2014}%
  \BibitemOpen
  \bibfield  {author} {\bibinfo {author} {\bibfnamefont {A.}~\bibnamefont
  {Tkatchenko}},\ }\href {\doibase 10.1002/adaaaaafm.201403029} {\bibfield
  {journal} {\bibinfo  {journal} {Advanced Functional Materials}\ }\textbf
  {\bibinfo {volume} {25}},\ \bibinfo {pages} {2054} (\bibinfo {year}
  {2015})}\BibitemShut {NoStop}%
\bibitem [{\citenamefont {Ambrosetti}\ \emph {et~al.}(2016)\citenamefont
  {Ambrosetti}, \citenamefont {Ferri}, \citenamefont {DiStasio},\ and\
  \citenamefont {Tkatchenko}}]{AmbrosettiSCIENCE2016}%
  \BibitemOpen
  \bibfield  {author} {\bibinfo {author} {\bibfnamefont {A.}~\bibnamefont
  {Ambrosetti}}, \bibinfo {author} {\bibfnamefont {N.}~\bibnamefont {Ferri}},
  \bibinfo {author} {\bibfnamefont {R.~A.}\ \bibnamefont {DiStasio},
  \bibfnamefont {Jr.}}, \ and\ \bibinfo {author} {\bibfnamefont
  {A.}~\bibnamefont {Tkatchenko}},\ }\href {\doibase 10.1126/science.aae0509}
  {\bibfield  {journal} {\bibinfo  {journal} {Science}\ }\textbf {\bibinfo
  {volume} {351}},\ \bibinfo {pages} {1171} (\bibinfo {year}
  {2016})}\BibitemShut {NoStop}%
\bibitem [{\citenamefont {Henry}\ and\ \citenamefont
  {Chen}(2008)}]{HenryPRL2008}%
  \BibitemOpen
  \bibfield  {author} {\bibinfo {author} {\bibfnamefont {A.}~\bibnamefont
  {Henry}}\ and\ \bibinfo {author} {\bibfnamefont {G.}~\bibnamefont {Chen}},\
  }\href {\doibase 10.1103/PhysRevLett.101.235502} {\bibfield  {journal}
  {\bibinfo  {journal} {Phys. Rev. Lett.}\ }\textbf {\bibinfo {volume} {101}},\
  \bibinfo {pages} {235502} (\bibinfo {year} {2008})}\BibitemShut {NoStop}%
\bibitem [{\citenamefont {Esfarjani}\ \emph {et~al.}(2011)\citenamefont
  {Esfarjani}, \citenamefont {Chen},\ and\ \citenamefont
  {Stokes}}]{EsfarjaniPRB2011}%
  \BibitemOpen
  \bibfield  {author} {\bibinfo {author} {\bibfnamefont {K.}~\bibnamefont
  {Esfarjani}}, \bibinfo {author} {\bibfnamefont {G.}~\bibnamefont {Chen}}, \
  and\ \bibinfo {author} {\bibfnamefont {H.~T.}\ \bibnamefont {Stokes}},\
  }\href {\doibase 10.1103/PhysRevB.84.085204} {\bibfield  {journal} {\bibinfo
  {journal} {Phys. Rev. B}\ }\textbf {\bibinfo {volume} {84}},\ \bibinfo
  {pages} {085204} (\bibinfo {year} {2011})}\BibitemShut {NoStop}%
\bibitem [{\citenamefont {Gonz\'alez~Noya}\ \emph {et~al.}(2004)\citenamefont
  {Gonz\'alez~Noya}, \citenamefont {Srivastava}, \citenamefont
  {Chernozatonskii},\ and\ \citenamefont {Menon}}]{NoyaPRB2004}%
  \BibitemOpen
  \bibfield  {author} {\bibinfo {author} {\bibfnamefont {E.}~\bibnamefont
  {Gonz\'alez~Noya}}, \bibinfo {author} {\bibfnamefont {D.}~\bibnamefont
  {Srivastava}}, \bibinfo {author} {\bibfnamefont {L.~A.}\ \bibnamefont
  {Chernozatonskii}}, \ and\ \bibinfo {author} {\bibfnamefont {M.}~\bibnamefont
  {Menon}},\ }\href {\doibase 10.1103/PhysRevB.70.115416} {\bibfield  {journal}
  {\bibinfo  {journal} {Phys. Rev. B}\ }\textbf {\bibinfo {volume} {70}},\
  \bibinfo {pages} {115416} (\bibinfo {year} {2004})}\BibitemShut {NoStop}%
\bibitem [{\citenamefont {Messina}\ \emph {et~al.}(2013)\citenamefont
  {Messina}, \citenamefont {Tschikin}, \citenamefont {Biehs},\ and\
  \citenamefont {Ben-Abdallah}}]{MessinaPRB2013}%
  \BibitemOpen
  \bibfield  {author} {\bibinfo {author} {\bibfnamefont {R.}~\bibnamefont
  {Messina}}, \bibinfo {author} {\bibfnamefont {M.}~\bibnamefont {Tschikin}},
  \bibinfo {author} {\bibfnamefont {S.-A.}\ \bibnamefont {Biehs}}, \ and\
  \bibinfo {author} {\bibfnamefont {P.}~\bibnamefont {Ben-Abdallah}},\ }\href
  {\doibase 10.1103/PhysRevB.88.104307} {\bibfield  {journal} {\bibinfo
  {journal} {Phys. Rev. B}\ }\textbf {\bibinfo {volume} {88}},\ \bibinfo
  {pages} {104307} (\bibinfo {year} {2013})}\BibitemShut {NoStop}%
\bibitem [{\citenamefont {P\'erez-Madrid}\ \emph {et~al.}(2008)\citenamefont
  {P\'erez-Madrid}, \citenamefont {Rub\'{\i}},\ and\ \citenamefont
  {Lapas}}]{PerezMadridPRB2008}%
  \BibitemOpen
  \bibfield  {author} {\bibinfo {author} {\bibfnamefont {A.}~\bibnamefont
  {P\'erez-Madrid}}, \bibinfo {author} {\bibfnamefont {J.~M.}\ \bibnamefont
  {Rub\'{\i}}}, \ and\ \bibinfo {author} {\bibfnamefont {L.~C.}\ \bibnamefont
  {Lapas}},\ }\href {\doibase 10.1103/PhysRevB.77.155417} {\bibfield  {journal}
  {\bibinfo  {journal} {Phys. Rev. B}\ }\textbf {\bibinfo {volume} {77}},\
  \bibinfo {pages} {155417} (\bibinfo {year} {2008})}\BibitemShut {NoStop}%
\bibitem [{\citenamefont {Edalatpour}\ and\ \citenamefont
  {Francoeur}(2014)}]{EdalatpourJQSRT2014}%
  \BibitemOpen
  \bibfield  {author} {\bibinfo {author} {\bibfnamefont {S.}~\bibnamefont
  {Edalatpour}}\ and\ \bibinfo {author} {\bibfnamefont {M.}~\bibnamefont
  {Francoeur}},\ }\href {\doibase https://doi.org/10.1016/j.jqsrt.2013.08.021}
  {\bibfield  {journal} {\bibinfo  {journal} {Journal of Quantitative
  Spectroscopy and Radiative Transfer}\ }\textbf {\bibinfo {volume} {133}},\
  \bibinfo {pages} {364} (\bibinfo {year} {2014})}\BibitemShut {NoStop}%
\bibitem [{\citenamefont {Edalatpour}\ \emph {et~al.}(2016)\citenamefont
  {Edalatpour}, \citenamefont {DeSutter},\ and\ \citenamefont
  {Francoeur}}]{EdalatpourJQSRT2016}%
  \BibitemOpen
  \bibfield  {author} {\bibinfo {author} {\bibfnamefont {S.}~\bibnamefont
  {Edalatpour}}, \bibinfo {author} {\bibfnamefont {J.}~\bibnamefont
  {DeSutter}}, \ and\ \bibinfo {author} {\bibfnamefont {M.}~\bibnamefont
  {Francoeur}},\ }\href {\doibase https://doi.org/10.1016/j.jqsrt.2015.12.027}
  {\bibfield  {journal} {\bibinfo  {journal} {Journal of Quantitative
  Spectroscopy and Radiative Transfer}\ }\textbf {\bibinfo {volume} {178}},\
  \bibinfo {pages} {14} (\bibinfo {year} {2016})}\BibitemShut {NoStop}%
\bibitem [{\citenamefont {Yu}\ \emph {et~al.}(2017)\citenamefont {Yu},
  \citenamefont {Manjavacas},\ and\ \citenamefont {de~Abajo}}]{YuNATURE2017}%
  \BibitemOpen
  \bibfield  {author} {\bibinfo {author} {\bibfnamefont {R.}~\bibnamefont
  {Yu}}, \bibinfo {author} {\bibfnamefont {A.}~\bibnamefont {Manjavacas}}, \
  and\ \bibinfo {author} {\bibfnamefont {F.~J.~G.}\ \bibnamefont {de~Abajo}},\
  }\href@noop {} {\bibfield  {journal} {\bibinfo  {journal} {Nature
  Communications}\ }\textbf {\bibinfo {volume} {8}},\ \bibinfo {pages} {2}
  (\bibinfo {year} {2017})}\BibitemShut {NoStop}%
\bibitem [{\citenamefont {Kr\"uger}\ \emph {et~al.}(2012)\citenamefont
  {Kr\"uger}, \citenamefont {Bimonte}, \citenamefont {Emig},\ and\
  \citenamefont {Kardar}}]{KrugerPRB2012}%
  \BibitemOpen
  \bibfield  {author} {\bibinfo {author} {\bibfnamefont {M.}~\bibnamefont
  {Kr\"uger}}, \bibinfo {author} {\bibfnamefont {G.}~\bibnamefont {Bimonte}},
  \bibinfo {author} {\bibfnamefont {T.}~\bibnamefont {Emig}}, \ and\ \bibinfo
  {author} {\bibfnamefont {M.}~\bibnamefont {Kardar}},\ }\href {\doibase
  10.1103/PhysRevB.86.115423} {\bibfield  {journal} {\bibinfo  {journal} {Phys.
  Rev. B}\ }\textbf {\bibinfo {volume} {86}},\ \bibinfo {pages} {115423}
  (\bibinfo {year} {2012})}\BibitemShut {NoStop}%
\bibitem [{\citenamefont {Molesky}\ and\ \citenamefont
  {Jacob}(2015)}]{MoleskyPRB2015}%
  \BibitemOpen
  \bibfield  {author} {\bibinfo {author} {\bibfnamefont {S.}~\bibnamefont
  {Molesky}}\ and\ \bibinfo {author} {\bibfnamefont {Z.}~\bibnamefont
  {Jacob}},\ }\href {\doibase 10.1103/PhysRevB.91.205435} {\bibfield  {journal}
  {\bibinfo  {journal} {Phys. Rev. B}\ }\textbf {\bibinfo {volume} {91}},\
  \bibinfo {pages} {205435} (\bibinfo {year} {2015})}\BibitemShut {NoStop}%
\bibitem [{\citenamefont {Zhao}\ \emph {et~al.}(2017)\citenamefont {Zhao},
  \citenamefont {Guizal}, \citenamefont {Zhang}, \citenamefont {Fan},\ and\
  \citenamefont {Antezza}}]{ZhaoPRB2017}%
  \BibitemOpen
  \bibfield  {author} {\bibinfo {author} {\bibfnamefont {B.}~\bibnamefont
  {Zhao}}, \bibinfo {author} {\bibfnamefont {B.}~\bibnamefont {Guizal}},
  \bibinfo {author} {\bibfnamefont {Z.~M.}\ \bibnamefont {Zhang}}, \bibinfo
  {author} {\bibfnamefont {S.}~\bibnamefont {Fan}}, \ and\ \bibinfo {author}
  {\bibfnamefont {M.}~\bibnamefont {Antezza}},\ }\href {\doibase
  10.1103/PhysRevB.95.245437} {\bibfield  {journal} {\bibinfo  {journal} {Phys.
  Rev. B}\ }\textbf {\bibinfo {volume} {95}},\ \bibinfo {pages} {245437}
  (\bibinfo {year} {2017})}\BibitemShut {NoStop}%
\bibitem [{\citenamefont {Otey}\ \emph {et~al.}(2014)\citenamefont {Otey},
  \citenamefont {Zhu}, \citenamefont {Sandhu},\ and\ \citenamefont
  {Fan}}]{OteyJQSRT2014}%
  \BibitemOpen
  \bibfield  {author} {\bibinfo {author} {\bibfnamefont {C.~R.}\ \bibnamefont
  {Otey}}, \bibinfo {author} {\bibfnamefont {L.}~\bibnamefont {Zhu}}, \bibinfo
  {author} {\bibfnamefont {S.}~\bibnamefont {Sandhu}}, \ and\ \bibinfo {author}
  {\bibfnamefont {S.}~\bibnamefont {Fan}},\ }\href {\doibase
  https://doi.org/10.1016/j.jqsrt.2013.04.017} {\bibfield  {journal} {\bibinfo
  {journal} {Journal of Quantitative Spectroscopy and Radiative Transfer}\
  }\textbf {\bibinfo {volume} {132}},\ \bibinfo {pages} {3} (\bibinfo {year}
  {2014})}\BibitemShut {NoStop}%
\bibitem [{\citenamefont {Reid}\ \emph {et~al.}(2013)\citenamefont {Reid},
  \citenamefont {Rodriguez},\ and\ \citenamefont {Johnson}}]{ReidPROCIEEE2013}%
  \BibitemOpen
  \bibfield  {author} {\bibinfo {author} {\bibfnamefont {M.~T.~H.}\
  \bibnamefont {Reid}}, \bibinfo {author} {\bibfnamefont {A.~W.}\ \bibnamefont
  {Rodriguez}}, \ and\ \bibinfo {author} {\bibfnamefont {S.~G.}\ \bibnamefont
  {Johnson}},\ }\href {\doibase 10.1109/JPROC.2012.2191749} {\bibfield
  {journal} {\bibinfo  {journal} {Proceedings of the IEEE}\ }\textbf {\bibinfo
  {volume} {101}},\ \bibinfo {pages} {531} (\bibinfo {year}
  {2013})}\BibitemShut {NoStop}%
\bibitem [{\citenamefont {Rodriguez}\ \emph {et~al.}(2011)\citenamefont
  {Rodriguez}, \citenamefont {Ilic}, \citenamefont {Bermel}, \citenamefont
  {Celanovic}, \citenamefont {Joannopoulos}, \citenamefont {Solja\ifmmode
  \check{c}\else \v{c}\fi{}i\ifmmode~\acute{c}\else \'{c}\fi{}},\ and\
  \citenamefont {Johnson}}]{RodriguezPRL2011}%
  \BibitemOpen
  \bibfield  {author} {\bibinfo {author} {\bibfnamefont {A.~W.}\ \bibnamefont
  {Rodriguez}}, \bibinfo {author} {\bibfnamefont {O.}~\bibnamefont {Ilic}},
  \bibinfo {author} {\bibfnamefont {P.}~\bibnamefont {Bermel}}, \bibinfo
  {author} {\bibfnamefont {I.}~\bibnamefont {Celanovic}}, \bibinfo {author}
  {\bibfnamefont {J.~D.}\ \bibnamefont {Joannopoulos}}, \bibinfo {author}
  {\bibfnamefont {M.}~\bibnamefont {Solja\ifmmode \check{c}\else
  \v{c}\fi{}i\ifmmode~\acute{c}\else \'{c}\fi{}}}, \ and\ \bibinfo {author}
  {\bibfnamefont {S.~G.}\ \bibnamefont {Johnson}},\ }\href {\doibase
  10.1103/PhysRevLett.107.114302} {\bibfield  {journal} {\bibinfo  {journal}
  {Phys. Rev. Lett.}\ }\textbf {\bibinfo {volume} {107}},\ \bibinfo {pages}
  {114302} (\bibinfo {year} {2011})}\BibitemShut {NoStop}%
\bibitem [{\citenamefont {Luo}\ \emph {et~al.}(2004)\citenamefont {Luo},
  \citenamefont {Narayanaswamy}, \citenamefont {Chen},\ and\ \citenamefont
  {Joannopoulos}}]{LuoPRL2004}%
  \BibitemOpen
  \bibfield  {author} {\bibinfo {author} {\bibfnamefont {C.}~\bibnamefont
  {Luo}}, \bibinfo {author} {\bibfnamefont {A.}~\bibnamefont {Narayanaswamy}},
  \bibinfo {author} {\bibfnamefont {G.}~\bibnamefont {Chen}}, \ and\ \bibinfo
  {author} {\bibfnamefont {J.~D.}\ \bibnamefont {Joannopoulos}},\ }\href
  {\doibase 10.1103/PhysRevLett.93.213905} {\bibfield  {journal} {\bibinfo
  {journal} {Phys. Rev. Lett.}\ }\textbf {\bibinfo {volume} {93}},\ \bibinfo
  {pages} {213905} (\bibinfo {year} {2004})}\BibitemShut {NoStop}%
\bibitem [{\citenamefont {Jablan}\ \emph {et~al.}(2011)\citenamefont {Jablan},
  \citenamefont {Solja\ifmmode \check{c}\else
  \v{c}\fi{}i\ifmmode~\acute{c}\else \'{c}\fi{}},\ and\ \citenamefont
  {Buljan}}]{JablanPRB2011}%
  \BibitemOpen
  \bibfield  {author} {\bibinfo {author} {\bibfnamefont {M.}~\bibnamefont
  {Jablan}}, \bibinfo {author} {\bibfnamefont {M.}~\bibnamefont {Solja\ifmmode
  \check{c}\else \v{c}\fi{}i\ifmmode~\acute{c}\else \'{c}\fi{}}}, \ and\
  \bibinfo {author} {\bibfnamefont {H.}~\bibnamefont {Buljan}},\ }\href
  {\doibase 10.1103/PhysRevB.83.161409} {\bibfield  {journal} {\bibinfo
  {journal} {Phys. Rev. B}\ }\textbf {\bibinfo {volume} {83}},\ \bibinfo
  {pages} {161409} (\bibinfo {year} {2011})}\BibitemShut {NoStop}%
\bibitem [{\citenamefont {Hwang}\ \emph {et~al.}(2010)\citenamefont {Hwang},
  \citenamefont {Sensarma},\ and\ \citenamefont {Das~Sarma}}]{HwangPRB2010}%
  \BibitemOpen
  \bibfield  {author} {\bibinfo {author} {\bibfnamefont {E.~H.}\ \bibnamefont
  {Hwang}}, \bibinfo {author} {\bibfnamefont {R.}~\bibnamefont {Sensarma}}, \
  and\ \bibinfo {author} {\bibfnamefont {S.}~\bibnamefont {Das~Sarma}},\ }\href
  {\doibase 10.1103/PhysRevB.82.195406} {\bibfield  {journal} {\bibinfo
  {journal} {Phys. Rev. B}\ }\textbf {\bibinfo {volume} {82}},\ \bibinfo
  {pages} {195406} (\bibinfo {year} {2010})}\BibitemShut {NoStop}%
\bibitem [{\citenamefont {Xiao}\ \emph {et~al.}(2016)\citenamefont {Xiao},
  \citenamefont {Zhu}, \citenamefont {Li},\ and\ \citenamefont
  {Mortensen}}]{XiaoFOP2016}%
  \BibitemOpen
  \bibfield  {author} {\bibinfo {author} {\bibfnamefont {S.}~\bibnamefont
  {Xiao}}, \bibinfo {author} {\bibfnamefont {X.}~\bibnamefont {Zhu}}, \bibinfo
  {author} {\bibfnamefont {B.-H.}\ \bibnamefont {Li}}, \ and\ \bibinfo {author}
  {\bibfnamefont {N.~A.}\ \bibnamefont {Mortensen}},\ }\href {\doibase
  10.1007/s11467-016-0551-z} {\bibfield  {journal} {\bibinfo  {journal}
  {Frontiers of Physics}\ }\textbf {\bibinfo {volume} {11}},\ \bibinfo {pages}
  {117801} (\bibinfo {year} {2016})}\BibitemShut {NoStop}%
\bibitem [{\citenamefont {Venkataram}\ \emph {et~al.}(2017)\citenamefont
  {Venkataram}, \citenamefont {Hermann}, \citenamefont {Tkatchenko},\ and\
  \citenamefont {Rodriguez}}]{VenkataramPRL2017}%
  \BibitemOpen
  \bibfield  {author} {\bibinfo {author} {\bibfnamefont {P.~S.}\ \bibnamefont
  {Venkataram}}, \bibinfo {author} {\bibfnamefont {J.}~\bibnamefont {Hermann}},
  \bibinfo {author} {\bibfnamefont {A.}~\bibnamefont {Tkatchenko}}, \ and\
  \bibinfo {author} {\bibfnamefont {A.~W.}\ \bibnamefont {Rodriguez}},\ }\href
  {\doibase 10.1103/PhysRevLett.118.266802} {\bibfield  {journal} {\bibinfo
  {journal} {Phys. Rev. Lett.}\ }\textbf {\bibinfo {volume} {118}},\ \bibinfo
  {pages} {266802} (\bibinfo {year} {2017})}\BibitemShut {NoStop}%
\bibitem [{\citenamefont {Song}\ \emph {et~al.}(2015)\citenamefont {Song},
  \citenamefont {Ganjeh}, \citenamefont {Sadat}, \citenamefont {Thompson},
  \citenamefont {Fiorino}, \citenamefont {Fern{\'a}ndez-Hurtado}, \citenamefont
  {Feist}, \citenamefont {Garcia-Vidal}, \citenamefont {Cuevas}, \citenamefont
  {Reddy} \emph {et~al.}}]{SongNATURENANO2015}%
  \BibitemOpen
  \bibfield  {author} {\bibinfo {author} {\bibfnamefont {B.}~\bibnamefont
  {Song}}, \bibinfo {author} {\bibfnamefont {Y.}~\bibnamefont {Ganjeh}},
  \bibinfo {author} {\bibfnamefont {S.}~\bibnamefont {Sadat}}, \bibinfo
  {author} {\bibfnamefont {D.}~\bibnamefont {Thompson}}, \bibinfo {author}
  {\bibfnamefont {A.}~\bibnamefont {Fiorino}}, \bibinfo {author} {\bibfnamefont
  {V.}~\bibnamefont {Fern{\'a}ndez-Hurtado}}, \bibinfo {author} {\bibfnamefont
  {J.}~\bibnamefont {Feist}}, \bibinfo {author} {\bibfnamefont {F.~J.}\
  \bibnamefont {Garcia-Vidal}}, \bibinfo {author} {\bibfnamefont {J.~C.}\
  \bibnamefont {Cuevas}}, \bibinfo {author} {\bibfnamefont {P.}~\bibnamefont
  {Reddy}},  \emph {et~al.},\ }\href@noop {} {\bibfield  {journal} {\bibinfo
  {journal} {Nature nanotechnology}\ }\textbf {\bibinfo {volume} {10}},\
  \bibinfo {pages} {253} (\bibinfo {year} {2015})}\BibitemShut {NoStop}%
\bibitem [{\citenamefont {Rahi}\ \emph {et~al.}(2009)\citenamefont {Rahi},
  \citenamefont {Emig}, \citenamefont {Graham}, \citenamefont {Jaffe},\ and\
  \citenamefont {Kardar}}]{RahiPRD2009}%
  \BibitemOpen
  \bibfield  {author} {\bibinfo {author} {\bibfnamefont {S.~J.}\ \bibnamefont
  {Rahi}}, \bibinfo {author} {\bibfnamefont {T.}~\bibnamefont {Emig}}, \bibinfo
  {author} {\bibfnamefont {N.}~\bibnamefont {Graham}}, \bibinfo {author}
  {\bibfnamefont {R.~L.}\ \bibnamefont {Jaffe}}, \ and\ \bibinfo {author}
  {\bibfnamefont {M.}~\bibnamefont {Kardar}},\ }\href@noop {} {\bibfield
  {journal} {\bibinfo  {journal} {Phys. Rev. D}\ }\textbf {\bibinfo {volume}
  {80}},\ \bibinfo {pages} {085021} (\bibinfo {year} {2009})}\BibitemShut
  {NoStop}%
\bibitem [{\citenamefont {Ambrosetti}\ \emph {et~al.}(2014)\citenamefont
  {Ambrosetti}, \citenamefont {Reilly}, \citenamefont {DiStasio},\ and\
  \citenamefont {Tkatchenko}}]{AmbrosettiJCP2014}%
  \BibitemOpen
  \bibfield  {author} {\bibinfo {author} {\bibfnamefont {A.}~\bibnamefont
  {Ambrosetti}}, \bibinfo {author} {\bibfnamefont {A.~M.}\ \bibnamefont
  {Reilly}}, \bibinfo {author} {\bibfnamefont {R.~A.}\ \bibnamefont
  {DiStasio}}, \ and\ \bibinfo {author} {\bibfnamefont {A.}~\bibnamefont
  {Tkatchenko}},\ }\href@noop {} {\bibfield  {journal} {\bibinfo  {journal}
  {The Journal of Chemical Physics}\ }\textbf {\bibinfo {volume} {140}}
  (\bibinfo {year} {2014})}\BibitemShut {NoStop}%
\bibitem [{\citenamefont {Volokitin}\ and\ \citenamefont
  {Persson}(2001)}]{VolokitinPRB2001}%
  \BibitemOpen
  \bibfield  {author} {\bibinfo {author} {\bibfnamefont {A.~I.}\ \bibnamefont
  {Volokitin}}\ and\ \bibinfo {author} {\bibfnamefont {B.~N.~J.}\ \bibnamefont
  {Persson}},\ }\href {\doibase 10.1103/PhysRevB.63.205404} {\bibfield
  {journal} {\bibinfo  {journal} {Phys. Rev. B}\ }\textbf {\bibinfo {volume}
  {63}},\ \bibinfo {pages} {205404} (\bibinfo {year} {2001})}\BibitemShut
  {NoStop}%
\bibitem [{\citenamefont {Domingues}\ \emph {et~al.}(2005)\citenamefont
  {Domingues}, \citenamefont {Volz}, \citenamefont {Joulain},\ and\
  \citenamefont {Greffet}}]{DominguesPRL2005}%
  \BibitemOpen
  \bibfield  {author} {\bibinfo {author} {\bibfnamefont {G.}~\bibnamefont
  {Domingues}}, \bibinfo {author} {\bibfnamefont {S.}~\bibnamefont {Volz}},
  \bibinfo {author} {\bibfnamefont {K.}~\bibnamefont {Joulain}}, \ and\
  \bibinfo {author} {\bibfnamefont {J.-J.}\ \bibnamefont {Greffet}},\ }\href
  {\doibase 10.1103/PhysRevLett.94.085901} {\bibfield  {journal} {\bibinfo
  {journal} {Phys. Rev. Lett.}\ }\textbf {\bibinfo {volume} {94}},\ \bibinfo
  {pages} {085901} (\bibinfo {year} {2005})}\BibitemShut {NoStop}%
\bibitem [{\citenamefont {Donchev}(2006)}]{DonchevJCP2006}%
  \BibitemOpen
  \bibfield  {author} {\bibinfo {author} {\bibfnamefont {A.~G.}\ \bibnamefont
  {Donchev}},\ }\href@noop {} {\bibfield  {journal} {\bibinfo  {journal} {The
  Journal of Chemical Physics}\ }\textbf {\bibinfo {volume} {125}} (\bibinfo
  {year} {2006})}\BibitemShut {NoStop}%
\bibitem [{\citenamefont {Phan}\ \emph {et~al.}(2013)\citenamefont {Phan},
  \citenamefont {Woods},\ and\ \citenamefont {Phan}}]{PhanJAP2013}%
  \BibitemOpen
  \bibfield  {author} {\bibinfo {author} {\bibfnamefont {A.~D.}\ \bibnamefont
  {Phan}}, \bibinfo {author} {\bibfnamefont {L.~M.}\ \bibnamefont {Woods}}, \
  and\ \bibinfo {author} {\bibfnamefont {T.-L.}\ \bibnamefont {Phan}},\ }\href
  {\doibase http://dx.doi.org/10.1063/1.4816446} {\bibfield  {journal}
  {\bibinfo  {journal} {Journal of Applied Physics}\ }\textbf {\bibinfo
  {volume} {114}} (\bibinfo {year} {2013}),\
  http://dx.doi.org/10.1063/1.4816446}\BibitemShut {NoStop}%
\bibitem [{\citenamefont {Shtogun}\ and\ \citenamefont
  {Woods}(2010)}]{ShtogunJPCL2010}%
  \BibitemOpen
  \bibfield  {author} {\bibinfo {author} {\bibfnamefont {Y.~V.}\ \bibnamefont
  {Shtogun}}\ and\ \bibinfo {author} {\bibfnamefont {L.~M.}\ \bibnamefont
  {Woods}},\ }\href@noop {} {\bibfield  {journal} {\bibinfo  {journal} {The
  Journal of Physical Chemistry Letters}\ }\textbf {\bibinfo {volume} {1}},\
  \bibinfo {pages} {1356} (\bibinfo {year} {2010})}\BibitemShut {NoStop}%
\bibitem [{\citenamefont {Kim}\ \emph {et~al.}(2007)\citenamefont {Kim},
  \citenamefont {Sofo}, \citenamefont {Velegol}, \citenamefont {Cole},\ and\
  \citenamefont {Lucas}}]{KimLANGMUIR2007}%
  \BibitemOpen
  \bibfield  {author} {\bibinfo {author} {\bibfnamefont {H.-Y.}\ \bibnamefont
  {Kim}}, \bibinfo {author} {\bibfnamefont {J.~O.}\ \bibnamefont {Sofo}},
  \bibinfo {author} {\bibfnamefont {D.}~\bibnamefont {Velegol}}, \bibinfo
  {author} {\bibfnamefont {M.~W.}\ \bibnamefont {Cole}}, \ and\ \bibinfo
  {author} {\bibfnamefont {A.~A.}\ \bibnamefont {Lucas}},\ }\href@noop {}
  {\bibfield  {journal} {\bibinfo  {journal} {Langmuir}\ }\textbf {\bibinfo
  {volume} {23}},\ \bibinfo {pages} {1735} (\bibinfo {year}
  {2007})}\BibitemShut {NoStop}%
\bibitem [{\citenamefont {Cole}\ \emph {et~al.}(2009)\citenamefont {Cole},
  \citenamefont {Velegol}, \citenamefont {Kim},\ and\ \citenamefont
  {Lucas}}]{ColeMS2009}%
  \BibitemOpen
  \bibfield  {author} {\bibinfo {author} {\bibfnamefont {M.~W.}\ \bibnamefont
  {Cole}}, \bibinfo {author} {\bibfnamefont {D.}~\bibnamefont {Velegol}},
  \bibinfo {author} {\bibfnamefont {H.-Y.}\ \bibnamefont {Kim}}, \ and\
  \bibinfo {author} {\bibfnamefont {A.~A.}\ \bibnamefont {Lucas}},\ }\href@noop
  {} {\bibfield  {journal} {\bibinfo  {journal} {Molecular Simulation}\
  }\textbf {\bibinfo {volume} {35}},\ \bibinfo {pages} {849} (\bibinfo {year}
  {2009})}\BibitemShut {NoStop}%
\bibitem [{\citenamefont {Mahanty}\ and\ \citenamefont
  {Ninham}(1975)}]{MahantyJCSFT1975}%
  \BibitemOpen
  \bibfield  {author} {\bibinfo {author} {\bibfnamefont {J.}~\bibnamefont
  {Mahanty}}\ and\ \bibinfo {author} {\bibfnamefont {B.~W.}\ \bibnamefont
  {Ninham}},\ }\href {\doibase 10.1039/F29757100119} {\bibfield  {journal}
  {\bibinfo  {journal} {J. Chem. Soc.{,} Faraday Trans. 2}\ }\textbf {\bibinfo
  {volume} {71}},\ \bibinfo {pages} {119} (\bibinfo {year} {1975})}\BibitemShut
  {NoStop}%
\bibitem [{\citenamefont {Renne}(1971)}]{RennePHYSICA1971A}%
  \BibitemOpen
  \bibfield  {author} {\bibinfo {author} {\bibfnamefont {M.~J.}\ \bibnamefont
  {Renne}},\ }\href {\doibase http://dx.doi.org/10.1016/0031-8914(71)90070-X}
  {\bibfield  {journal} {\bibinfo  {journal} {Physica}\ }\textbf {\bibinfo
  {volume} {53}},\ \bibinfo {pages} {193} (\bibinfo {year} {1971})}\BibitemShut
  {NoStop}%
\bibitem [{\citenamefont {Mayer}(2007)}]{MayerPRB2007}%
  \BibitemOpen
  \bibfield  {author} {\bibinfo {author} {\bibfnamefont {A.}~\bibnamefont
  {Mayer}},\ }\href@noop {} {\bibfield  {journal} {\bibinfo  {journal} {Phys.
  Rev. B}\ }\textbf {\bibinfo {volume} {75}},\ \bibinfo {pages} {045407}
  (\bibinfo {year} {2007})}\BibitemShut {NoStop}%
\bibitem [{Note1()}]{Note1}%
  \BibitemOpen
  \bibinfo {note} {For computational convenience, and since gold is essentially
  perfectly reflecting in the infrared frequency regime where nonequilibrium
  energy transfer is most significant, we model the gold plate as a perfect
  electrically conducting surface}\BibitemShut {NoStop}%
\bibitem [{\citenamefont {Buhmann}\ \emph {et~al.}(2012)\citenamefont
  {Buhmann}, \citenamefont {Scheel}, \citenamefont {Ellingsen}, \citenamefont
  {Hornberger},\ and\ \citenamefont {Jacob}}]{BuhmannPRA2012}%
  \BibitemOpen
  \bibfield  {author} {\bibinfo {author} {\bibfnamefont {S.~Y.}\ \bibnamefont
  {Buhmann}}, \bibinfo {author} {\bibfnamefont {S.}~\bibnamefont {Scheel}},
  \bibinfo {author} {\bibfnamefont {S.~A.}\ \bibnamefont {Ellingsen}}, \bibinfo
  {author} {\bibfnamefont {K.}~\bibnamefont {Hornberger}}, \ and\ \bibinfo
  {author} {\bibfnamefont {A.}~\bibnamefont {Jacob}},\ }\href {\doibase
  10.1103/PhysRevA.85.042513} {\bibfield  {journal} {\bibinfo  {journal} {Phys.
  Rev. A}\ }\textbf {\bibinfo {volume} {85}},\ \bibinfo {pages} {042513}
  (\bibinfo {year} {2012})}\BibitemShut {NoStop}%
\bibitem [{\citenamefont {Axilrod}\ and\ \citenamefont
  {Teller}(1943)}]{AxilrodJCP1943}%
  \BibitemOpen
  \bibfield  {author} {\bibinfo {author} {\bibfnamefont {B.~M.}\ \bibnamefont
  {Axilrod}}\ and\ \bibinfo {author} {\bibfnamefont {E.}~\bibnamefont
  {Teller}},\ }\href {\doibase 10.1063/1.1723844} {\bibfield  {journal}
  {\bibinfo  {journal} {The Journal of Chemical Physics}\ }\textbf {\bibinfo
  {volume} {11}},\ \bibinfo {pages} {299} (\bibinfo {year} {1943})},\ \Eprint
  {http://arxiv.org/abs/https://doi.org/10.1063/1.1723844}
  {https://doi.org/10.1063/1.1723844} \BibitemShut {NoStop}%
\bibitem [{Note2()}]{Note2}%
  \BibitemOpen
  \bibinfo {note} {Note that in principle, as $d$ attains angstrom length
  scales, the assumption of the molecules being separate breaks down due to the
  possible formation of new covalent bonds; while this would require
  recomputing $K_{\protect \mathrm {I}}$ for the coupled molecular system, our
  formalism for computing radiative heat transfer still holds as long as the
  two molecular components can be maintained at different
  temperatures.}\BibitemShut {Stop}%
\bibitem [{Note3()}]{Note3}%
  \BibitemOpen
  \bibinfo {note} {Note that in principle, the response of the gold plate at
  nanometric separations should also be modified to account for atomic-scale
  effects, e.g. charge smearing, either phenomenologically by introducing a
  macroscopic nonlocal susceptibility (as is typically done~\cite
  {ChapuisPRB2008,SingerJQSRT2015,SchmidtJPB2018}) or atomistically within the
  proposed framework, but we leave this as subject of future work.}\BibitemShut
  {Stop}%
\bibitem [{\citenamefont {Chapuis}\ \emph {et~al.}(2008)\citenamefont
  {Chapuis}, \citenamefont {Volz}, \citenamefont {Henkel}, \citenamefont
  {Joulain},\ and\ \citenamefont {Greffet}}]{ChapuisPRB2008}%
  \BibitemOpen
  \bibfield  {author} {\bibinfo {author} {\bibfnamefont {P.-O.}\ \bibnamefont
  {Chapuis}}, \bibinfo {author} {\bibfnamefont {S.}~\bibnamefont {Volz}},
  \bibinfo {author} {\bibfnamefont {C.}~\bibnamefont {Henkel}}, \bibinfo
  {author} {\bibfnamefont {K.}~\bibnamefont {Joulain}}, \ and\ \bibinfo
  {author} {\bibfnamefont {J.-J.}\ \bibnamefont {Greffet}},\ }\href {\doibase
  10.1103/PhysRevB.77.035431} {\bibfield  {journal} {\bibinfo  {journal} {Phys.
  Rev. B}\ }\textbf {\bibinfo {volume} {77}},\ \bibinfo {pages} {035431}
  (\bibinfo {year} {2008})}\BibitemShut {NoStop}%
\bibitem [{\citenamefont {Singer}\ \emph {et~al.}(2015)\citenamefont {Singer},
  \citenamefont {Ezzahri},\ and\ \citenamefont {Joulain}}]{SingerJQSRT2015}%
  \BibitemOpen
  \bibfield  {author} {\bibinfo {author} {\bibfnamefont {F.}~\bibnamefont
  {Singer}}, \bibinfo {author} {\bibfnamefont {Y.}~\bibnamefont {Ezzahri}}, \
  and\ \bibinfo {author} {\bibfnamefont {K.}~\bibnamefont {Joulain}},\ }\href
  {\doibase https://doi.org/10.1016/j.jqsrt.2014.11.016} {\bibfield  {journal}
  {\bibinfo  {journal} {Journal of Quantitative Spectroscopy and Radiative
  Transfer}\ }\textbf {\bibinfo {volume} {154}},\ \bibinfo {pages} {55}
  (\bibinfo {year} {2015})}\BibitemShut {NoStop}%
\bibitem [{\citenamefont {Schmidt}\ and\ \citenamefont
  {Scheel}(2018)}]{SchmidtJPB2018}%
  \BibitemOpen
  \bibfield  {author} {\bibinfo {author} {\bibfnamefont {R.}~\bibnamefont
  {Schmidt}}\ and\ \bibinfo {author} {\bibfnamefont {S.}~\bibnamefont
  {Scheel}},\ }\href {http://stacks.iop.org/0953-4075/51/i=4/a=044003}
  {\bibfield  {journal} {\bibinfo  {journal} {Journal of Physics B: Atomic,
  Molecular and Optical Physics}\ }\textbf {\bibinfo {volume} {51}},\ \bibinfo
  {pages} {044003} (\bibinfo {year} {2018})}\BibitemShut {NoStop}%
\bibitem [{Nov(2006)}]{Novotny2006}%
  \BibitemOpen
  in\ \href@noop {} {\emph {\bibinfo {booktitle} {Principles of Nano-Optics}}}\
  (\bibinfo  {publisher} {Cambridge University Press},\ \bibinfo {year}
  {2006})\ pp.\ \bibinfo {pages} {335--362}\BibitemShut {NoStop}%
\end{thebibliography}%
\end{document}